\documentclass[arxiv, usenatbib]{iupartex}
\usepackage{lineno}
\usepackage{newtxtext,newtxmath}
\usepackage[T1]{fontenc}
\usepackage{ae,aecompl}

\usepackage{multicol}   
\usepackage{bm}		    
\usepackage{pdflscape}	

\title[Revisiting Bolometric Corrections]{Fundamentals of Stars II: Revisiting Bolometric Corrections}

\author[Eker \& Bak{\i}\c{s}]{%
Z.~Eker$^{1\cc}$\orcid{0000-0003-1883-6255},
and
V. Bak{\i}\c{s}$^{1}$\orcid{0000-0002-3125-9010}
\affsep \\
$^1$Akdeniz University, Faculty of Sciences, Department of Space Sciences and Technologies, 07058, Antalya, Türkiye\\
}

\corres{Z. Eker}{eker@akdeniz.edu.tr}

\pubyear{2025}

\doiheader{XXXXXXX/PAR.2025.00000}
\date{
	\pSubmit{21.02.2025} 
	\pAccept{19.03.2025}
	\pPubOnl{XX.XX.XXXX}
}
\volume{3}
\volnumber{1}

\begin{document}
\label{firstpage}
\pagerange{\pageref*{firstpage}--\pageref*{lastpage}}
 \maketitle

\begin{abstract}
 The development line of bolometric corrections within the brief history of photometry was described from the perspective of the Kuhnian philosophy of science. The luminous efficiency and heat index were two previous concepts to imply visual and bolometric brightness difference of a star, which was mainly suggested and used as auxiliary tools for calibrating stellar temperature scales before the term ``bolometric correction'' ({\it BC}) was also introduced for the same purpose by Kuiper in 1938, as $BC = M_{\rm bol} - M_{\rm V} = m_{\rm {bol}} - V$. Despite its ill-posed nature imposing no zero-point constant ($C_2=0$) for the {\it BC} scale and $L_{\rm V} = L \times 10^{BC/2.5}$, if $BC>0$, $L_{\rm V}$ is unphysical, for the luminosity of a star from which ``{\it BC} of a star must always be negative,'' ``the bolometric magnitude of a star ought to be brighter than its $V$-magnitude,'' and ``the zero point of bolometric corrections are arbitrary'' (paradigms) were extracted. The newest of the first three definitions of {\it BC} was accepted and used throughout the century. Therefore, the part of the development line of {\it BC} before Kuiper could be considered a prescience period. The rest could be named the normal science period in which astrophysicists work under the three paradigms. The rise of {\it BC} as a concept, how the ill-posed definition {\it BC} emerged/used, how inconsistencies (paradigms) of {\it BC} developed, and how the Resolution B2 of the General Assembly of the International Astronomical Union imposing $C_{\rm bol} = 71.197\,425\,\ldots$ mag, and $C_2>0$, for the zero-point constants of the $M_{\rm {Bol}}$ and {\it BC} scales resolve the long-lasting problems were discussed. Generalized new definition of {\it BC} implying $L_{\rm V}=L \times 10^\frac{({\rm BC}-C_2)}{2.5}$  were given to replace $L_{\rm V} = L \times 10^{BC/2.5}$.
\end{abstract}




\section{Introduction}
\label{sec:Introduction}

The stellar magnitude scale is well established today that with available technology, it is possible to measure as small as 0.3 mmag variation after {\it Gaia} with systematic errors less than one per cent, which means 0.028 per cent of brightness change of a star is detectable \citep{GaiaEDR3}. The first star catalogue containing 850 naked eye stars with coordinates and a simple brightness classification, ``of brilliant light'', ``of second degree'', or ``faint'', which was introduced by Hipparchus of Nicaea (190-120 BC), is the very first step of advancements in astronomical photometry \citep{Miles2007}. The original work of Hipparchus was lost \citep{Zissell1998}. Thus, our knowledge about it comes mainly from Almagest of Claudius Ptolemy (100-170 AD), who increased the number of stars to 1028 in his list for an epoch of about 137 AD, almost three centuries later. Ptolemy was the one who classified naked eye stars into six brightness classes, which is still used today, where the brightest stars were the first class, while the dimmest ones were in the sixth \citep{Miles2007}. 

In these early times, magnitudes were not a continuous real-number scale as it is used today, but just classes or bins of the observable stars classified in. Ptolemy's work and the magnitudes were revised later by Al-Sufi (903-986), who increased the number of stars to 1151, where Ptolemy’s positions precessed to the year 964 AD \citep{Zissell1998, Miles2007}. Al-Sufi's magnitudes were then revised by Ulugh Bey (1394-1449) into a system that indicates a rough position of a star within a magnitude class, e.g. 3, 4, 3-4, 4-3, 4-5 but still not a continuous scale. The brightness of 788 stars was listed in one-third magnitude steps in the Catalog of Tycho Brahe (1546-1601). This is because decimal points have not yet been invented \citep{Zissell1998}.

The discovery of the optical telescope led Galileo Galilei (1564-1642) to discover stars dimmer than six magnitudes. Nonetheless, Galilei did not attempt to quantify them. Edmond Halley (1656-1742) and John Flamsteed (1646-1719) were known to estimate stellar magnitudes using large sextants. Those early estimates, however, which could be in error up to 1.5 mag, thus, cannot properly be counted as an attempt at visual photometry, but rather as an aid to the stellar identifications \citep{Miles2007}. A strong need for a continuous magnitude scale was felt by William Herschel (1738-1822) for two reasons: First of all, Herschel started a project by a technique he called star-gauging to investigate Newton’s (1643-1727) claim that stars distributed uniformly in an endless universe in order not to collapse because of universal gravity. Assuming all stars have the same absolute brightness, e.g. being like the Sun, and dimming occurs due to inverse square law, more distant stars must appear fainter and fainter proportional to their distances. So, Herschell thought it was possible to determine how the stars distributed on a line of sight by counting and comparing the number of one magnitude dimmer stars to the number of stars one magnitude brighter continuously in the field of a telescope. Working together with his sister Caroline, he made star-gauging towards 700 different directions in the sky and determined the disk-like shape of the Milky Way, imagined as the universe at that time. This is because the space density of stars was observed to be decreasing by going further out, and there were no stars after certain distances \citep{Mihalas1981}. Thus, Herschel was interested in better brightness estimates than Halley and Flamsteed for his own sake. Secondly, and more importantly, Herschell was interested in determining the change in brightness of some stars, which were already known to vary and to discover whether other such variables existed \citep{Miles2007}. 

Soon after the invention of the visual photometer, the impulse initiated by Herschel gave its first fruit by Norman Robert Pogson, who formalized the magnitude system in 1856 by defining a first magnitude star as that is 100 times brighter than the sixth magnitude star \citep{Zissell1998, Cox2000}. Let us leave the details of how visual estimates were done, how visual photometers were constructed and worked, how catalogues such as Henry Draper (HD), Bonner Durchmusterung (BD) initiated to \citet{Zissell1998}, and \citet{Miles2007} and the references therein. Here, in this article, we are mostly interested in investigating when, why and how the other kinds of magnitudes, e.g. bolometric, photographic, blue, red, etc., first appeared in the history of photometry. What were the practical purposes and difficulties in determining and using bolometric corrections? When do paradigms such as ``the bolometric magnitude of a star ought to be brighter than its visual magnitude'', ``bolometric corrections must always be negative'' and ``the zero point of bolometric corrections are arbitrary'' entered in literature? How does the resolution of those problems benefit contemporary astrophysics?

\section{Emergence of Photographic and Bolometric Magnitudes}
\label{sec:Emergence of Photographic and Bolometric Magnitudes}

Another milestone on the progress line of photometry is the usage of photographic plates. Despite the first photographs dating back to 1826, more than 30 thirty years had to pass for early applications by wet collodion plates and at least another 30 years more to pass for the first successful applications by the dry plates that Revd Thomas Espin compiled 500 photographic magnitudes \citep{Miles2007}. Despite complexities caused by several other factors such as seeing, edge effects, plate-to-plate differences and so on, astronomical photometry was the forerunner of developments in astronomy during the late 19th and early 20th century because of its relative easiness with respect to visual photometry. In fact, it was easier and more reliable to measure the relative brightness of stars from their trailed images or by comparing the sizes of star images on an exposed plate. Jacobus Cornelius Kapteyn (1851-1922) and Henrietta Swan Leavitt (1968-1922) were two outstanding pioneers who used early photographic magnitudes. Kapteyn continued Herschel's star gauge with more reliable magnitudes from the catalogue of 454 000 stars in the Cape Photographic DM, where the image diameter was used as a function of magnitude, produced in the years 1885-1890 \citep{Zissell1998}. Star-gauging studies were then successful with the first Galaxy model in which the Milky Way was an oblate star system approximately 8.5 kpc in width and 1.7 kpc in thickness, having the Sun at about the centre \citep{Mihalas1981}. Henrietta Leavitt, on the other hand, became world-famous for discovering the period-luminosity relation of cepheid variables, which led to the establishment of a distance scale to nearby galaxies.    

A new kind of magnitude other than visual was first noticed because of photographic plates, such as the ones used by Kapteyn \citep{Zissell1998}. This is because it has been noticed that blue stars appeared brighter, while red stars appeared fainter on exposed plates in comparison to visual observations. The difference in the spectral sensitivity of photographic emulsions with respect to human eyes was a major issue. Filters had to be used until special emulsions imitating visual response were found. Blue-sensitive photographic magnitudes derived from unfiltered exposures existed side by side with the visual/photovisual magnitudes for many decades. What appeared as chaos turned out to be another step forward later in the line of development. That is the magnitude difference between a photographic magnitude $m_{\rm pg}$ and a visual $m_{\rm v}$ or photovisual $m_{\rm pv}$ is defined to be $(b-v)$ colour index, which was an important parameter to construct early colour-magnitude diagrams of nearby open clusters, e.g. Pleiades, and useful to understand stellar evolution, equivalent to the $B-V$ colour index today. After the H-R diagrams constructed in 1911 by \citet{Hertzsprung1911} and in 1913 by \citet{Russell1914} independently for the first time, the year 1913 was just the correct time for the Pogson scale of intensity ratio to be formally adopted to photographic photometry. Consequently, the zero point of the colour index was defined so that a star with a spectral type A0 should have equal visual and photographic magnitudes \citep{Miles2007}.

Although there were some careful observers of eclipsing binaries, such as \citet{Stebbins1910}, achieved probable errors of $\pm$0.023 mag near the principal minimum, and even milli magnitude level ($\pm$0.006 mag) of probable errors near the secondary minimum and between the minima by selenium photometers, \citet{Miles2007} estimated the systematic errors of the photographic system in the beginnings of the 20th century was about 0.2-0.3 mag and occasionally 0.5 mag, especially for the stars in the southern hemisphere, where transfer of magnitudes from NPS (North Polar Sequence) system was problematic. The problems were not only these but also those that some observers were choosing different stars for the zero-point of the magnitude scale in general. \citet{Eddington1926} chose the brighter component of Capella as the basis for theoretical bolometric magnitudes, but uncertainties in the relative brightness of the components of Capella and its parallax made this star unacceptable as the zero point \citep{Pettit1928}. Once, North Star Polaris, which was initially assumed to have a magnitude of 2, was tried to be used as the reference star, but later it was found to be a variable star at 2.12 mag with about four-day variations up to 0.17 mag and therefore found not suitable for this task \citep{Zissell1998}. Today, $\alpha$ Lyr (Vega) is commonly accepted to be the primary calibrating star for observable magnitudes and colours.

While astronomers who lived in the late 19th and early 20th centuries were busy with photographic plates, which led them to invent a new kind of magnitude, symbolized $m_{\rm pg}$, in addition to the visual magnitudes ($m_{\rm v}$ or $m_{\rm pv}$), the contemporary physicists working on radiometers, bolometers and black body radiation was on the way of introducing bolometric magnitudes. After the inventions of the radiometer \citep{Crookes1874}, the bolometer \citep{Langley1880}, and numerous theoretical attempts associated with laboratory experiments on the radiation laws of Kirchhoff, Rayleigh-Jeans and Wien, the radiation of a black body at all frequencies as a function of absolute temperature was finally expressed by \citet{Planck1900}. 

Although stars were recognized as radiating black bodies at different temperatures and sizes, the difference in the observed stellar SED from the Planck radiation law was often emphasized \citep{Eddington1926, Kuiper1938a}. \citet{Kuiper1938a} did not hesitate to say, ``It is well known that the stars do not radiate as black bodies'' while \citet{Eddington1926} wrote: The radiation of a star is not distributed exactly in accordance with the law of black-body radiation. If $T_{\rm e}$ stands for the effective temperature corresponding to the quantity of the radiant energy according to the Plank Law, and if \(T_{\rm e}'\) is the effective temperature obtained from the Wien Law; \(\lambda_{\text{max}} \cdot T_{\rm e}' = 0.288 \, \text{cm} ~\text{K}\), the general effect is that the quality of the radiation corresponds to a rather higher effective temperature than the quantity. ``For the Sun, \(T_{\rm e}'\) is about 4\% higher than \(T_{\rm e}\) (approximately \(6000^\circ\) against \(5740^\circ\)) and the same ratio may be expected to hold for all stars, at any rate as a first approximation''. Furthermore, the following comment of \citet{Eddington1926} is more interesting and crucial, especially for understanding why bolometric magnitudes were defined:

\begin{pquote}
It is now practical to measure the heat received from a star by the use of a radiometer. Considerable sensitiveness in the method has been developed (* It is said that the equipment at Mount Wilson could detect the heat of a candle on the banks of the Mississippi). But the results attained are as yet very limited, and in general, we have to infer the total amount of heat emitted from the light emitted (+ the deduction of bolometric magnitude from heat measurement is not really more direct than from light measurement, because large corrections must be applied on account of atmospheric absorption in the infra-red, and this involves assuming a SED just as the reduction of light measurements does). This involves the luminous efficiency of the energy emitted by stars of different types.
\end{pquote}

Readers should pay extra attention to the two issues in the above comment. The first is that the bolometric magnitude of a star $m_{\rm bol}$ was described as a directly measurable quantity such as $m_{\rm pg}$ or $m_{\rm v}$. This is certainly different from the present-day common sense impose $m_{\rm bol}$ as an unobservable quantity representing the total radiant energy from a star, called luminosity $L$. This is because there is no telescope or detector to operate at all wavelengths from zero to infinity. The second is that the bolometric magnitudes come from heat measurements while $m_{\rm pg}$ or $m_{\rm v}$ comes from light measurements as if heat and light were two different entities. The keyword is ``luminous efficiency'', which appears to be the first concept on the way of developing the final concept suggested by \citet{Kuiper1938a}, who named it ``bolometric correction'' to indicate the brightness difference of a star between its bolometric and visual magnitudes. 

Nevertheless, a new kind of magnitude other than photographic and visual entered the astronomical literature before the concept ``bolometric correction'' was introduced at the beginning of the 20th Century. There were only three kinds of magnitudes (bolometric, $b$ and $v$) until \citet{Johnson1953} who introduced the first standard three-colour photometry, where {\it UBV} in which ultraviolet magnitude $U$ could be considered newest and 4th kind of magnitude because other blue and yellow ($B$ and $V$) were there to meet the previous $b$ and $v$ magnitudes.

\section{Establishment of the Concept of Bolometric Correction}
\subsection{The Earliest Concept: Luminous Efficiency}

Before trying to understand why \citet{Eddington1926} advocated bolometric magnitude as a detectable rather than hypothetical brightness to indicate the total luminosity of a star, the following facts must be remembered: 1) F. W. Anton made precise measurements of the masses of many different atoms, including hydrogen and helium, by the year 1920. Anton’s results inspired Eddington to discover a fusion of hydrogen in the cores of stars \citep{Bahcall2000}, but neutrons were not yet discovered by \citet{Chadwick1933}. Therefore, CNO cycle reactions were not yet written by Hans Bethe and Von Weizsacher \citep{Clayton1968}, and the time was too early for \citet{Bethe1939} to establish p-p reactions. The stellar structure and evolution theory with nuclear reactions had not yet existed for Eddington. 2) Atmospheric extinction was known, and corrections due to Earth’s atmosphere were applied, but interstellar extinctions in most cases were not even mentioned before \citet{Trumpler1930}. 3) the famous main-sequence mass-luminosity relation (MLR) was discovered just recently in its primitive form called mass-visual absolute brightness relation by \citet{Hertzsprung1923} and \citet{Russell1923} independently, and \citet{Eddington1926} upgraded it as one of the fundamental laws of nature, like Planck's law or Stefan-Boltzmann law \citep{Eker2024}.

The joint discovery inspired Eddington for a more fundamental relation between the mass of a star and its luminosity because among the observable magnitudes of the time ($m_{\rm bol}$, $m_{\rm pg}$ and $m_{\rm pv}$) $m_{\rm bol}$ appeared to be the most suitable one to be associated with the luminosity of a star. This is also because heat measurements were believed to be directly proportional to the luminosity of a star despite the fact it must also be corrected for the atmospheric extinction in a similar manner $m_{\rm pg}$ and $m_{\rm pv}$ were done. Eddington first established tabulated numbers for a mass–absolute bolometric magnitude (Table \ref{tab:tab1}) relation assuming stars are hot gases of spheres using,
\begin{equation}
\label{eq:01}
\Delta m = -\frac{7}{2} \Delta \log M - \frac{15}{4} \Delta \log (1-\beta) - 2 \Delta \log T_{\rm e},
\end{equation}
where $M$ stands for mass, $(1-\beta)$ is the ratio of the radiation pressure to the whole pressure, $T_{\rm e}$ is the effective temperature corresponding to the quantity of the radiant energy according to the Planck’s Law, and $m$ is the bolometric magnitude. The mass-absolute bolometric magnitude ($M-M_{\rm bol}$) relation would have been beneficial to determine the masses of single stars since there were no other ways to obtain masses of single stars at these times.

\begin{table*}[h!]
\renewcommand{\tabcolsep}{4.0mm}
\renewcommand{\arraystretch}{1.1}
\centering
\caption{Data for the theoretical mass $M$ and absolute bolometric magnitude $m$ relation of \citet{Eddington1926} ($\mu = 2.11$, $T_{\rm e}=5200^\circ$).}
\begin{tabular}{|c|c|c|c|c|c|c|c|c|}
\hline
$1-\beta$ & $M$    & $m$     & $1-\beta$ & $M$   & $m$  & $1-\beta$ & $M$    & $m$           \\ \hline
0.0010  & 0.1284 & 14.143  & 0.04  & 0.879 & 5.211 & 0.26 & 3.774 & -0.052   \\ 
0.0015  & 0.1574 & 13.173  & 0.05  & 1.004 & 4.645 & 0.28 & 4.137 & -0.312   \\ 
0.0020  & 0.1820 & 12.484  & 0.06  & 1.123 & 4.178 & 0.30 & 4.529 & -0.562   \\ 
0.0025  & 0.2036 & 11.950  & 0.07  & 1.240 & 3.777 & 0.35 & 5.675 & -1.156   \\ 
0.0030  & 0.2233 & 11.513  & 0.08  & 1.354 & 3.426 & 0.40 & 7.117 & -1.718   \\ 
0.0040  & 0.2583 & 10.823  & 0.09  & 1.468 & 3.111 & 0.45 & 8.984 & -2.264   \\
0.0050  & 0.2895 & 10.286  & 0.10  & 1.582 & 2.825 & 0.50 & 11.46 & -2.805   \\
0.0060  & 0.3176 & ~~9.848 & 0.12  & 1.812 & 2.322 & 0.55 & 14.84 & -3.354   \\
0.0080  & 0.3683 & ~~9.154 & 0.14  & 2.050 & 1.884 & 0.60 & 19.62 & -3.919   \\
0.0100  & 0.4135 & ~~8.615 & 0.16  & 2.297 & 1.494 & 0.65 & 26.66 & -4.516   \\
0.0150  & 0.5117 & ~~7.632 & 0.18  & 2.557 & 1.138 & 0.70 & 37.67 & -5.162   \\
0.0200  & 0.5968 & ~~6.929 & 0.20  & 2.831 & 0.812 & 0.75 & 56.15 & -5.882   \\
0.0250  & 0.6739 & ~~6.381 & 0.22  & 3.124 & 0.507 & 0.80 & 90.63 & -6.714   \\
0.0300  & 0.7460 & ~~5.929 & 0.24  & 3.437 & 0.220 & ---  & ---   & ---      \\
\hline
\end{tabular}
\\
Add to $m$ temperature term, $-2\log_{\rm 10}=(T_{\rm e}/5200)$. 
\label{tab:tab1}
\end{table*}

\begin{figure}
\centering
\includegraphics[width=0.4\linewidth]{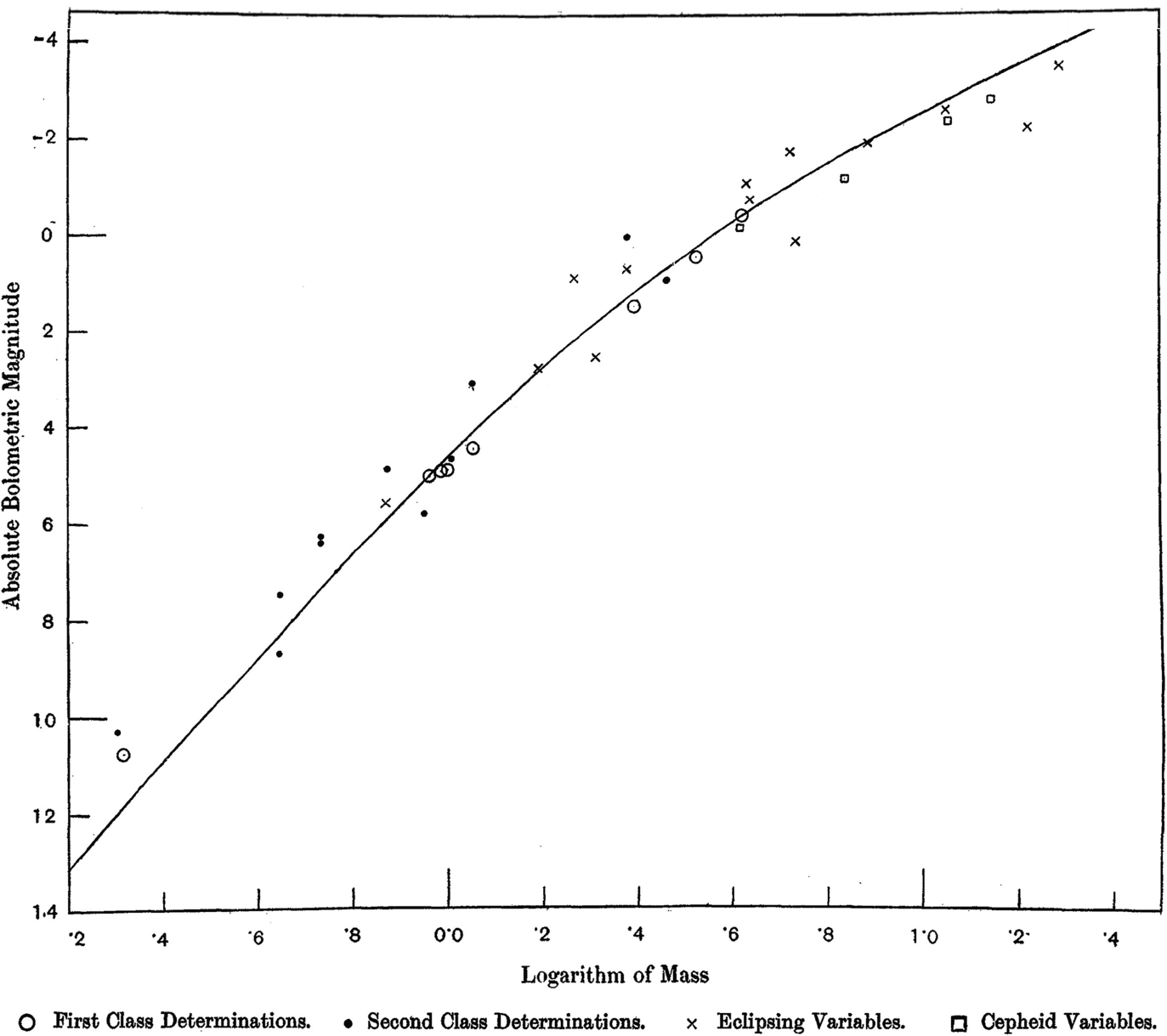}
\caption{Mass-Absolute Bolometric Magnitude curve of \citet{Eddington1926} according to the Equation~ (\ref{eq:01}) and stellar masses (data) available at that time. ``O'' are the masses of ordinary binaries of first-class determinations and Sun, filled circles are masses from ordinary binaries of second-class determinations, ``$X$'' are eclipsing variables, and ``$\square$'' cepheid variables} 
\label{fig:01}
\end {figure}

Column $m$ of Table \ref{tab:tab1} was computed by omitting the term with $T_{\rm e}$ in Equation~(\ref{eq:01}). The zero-point of bolometric absolute magnitudes was adjusted using the observed mass and absolute bolometric magnitude of Capella, which was known to have $T_{\rm e} = 5200^\mathrm{o}$ and chemical composition parameter $\mu = 2.11$. If a star has a different effective temperature than Capella ($5200^\mathrm{o}$), the term $\Delta m = -2 \Delta \log T_{\rm e}$ must be added to the m as indicated at the foot of the table. \citet{Eddington1926} claimed, ``With the range of effective temperature commonly occurring in stars, this correction is comparatively small, and the bolometric magnitude of a star is mainly a function of its mass.'' \citet{Eddington1926} was not satisfied only with the theoretical formula and the table of data because he also had an opportunity to confirm his claim by the observations of that time, as shown in Figure \ref{fig:01}.

Soon after, Eddington noticed that some stars have absolute bolometric magnitudes different than their absolute visual magnitudes, which could be interpreted as a deficit in the visual brightness of a star on the magnitude scale. Why hotter and cooler stars both appear dimmer in the visual was a problem of time which could not be solved solely by Planck’s law. Eddington managed the problem by suggesting a new concept, ``\textit{luminous efficiency}'' of the energy emitted by the stars of different types.

If a star radiates as a black body of temperature $T_{\rm e}$, the amount of radiation is $I'(\lambda, T_{\rm e}) d\lambda$ within the wavelength range from $\lambda$ to $\lambda + d\lambda$. Let $P(\lambda)$ be the factor that, if multiplied by the amount of radiation at wavelength $\lambda$, one can get its luminous intensity. One can assume $P(\lambda)$ is like the transparency of a filter so that it is zero for the wavelengths human eyes cannot see; otherwise, it has a positive value. The average of $P(\lambda)$ for the whole radiation is then given by:
\begin{equation}
\label{eq:02}
P = \frac{\int P(\lambda) I'(\lambda, T_{\rm e}) d\lambda}{\int I'(\lambda, T_{\rm e}) d\lambda}. 
\end{equation}

According to Eddington, the maximum of $P$ occurs at $T_{\rm e}=6 500^\mathrm{o}$  so that the stars of spectral types F and G have the greatest luminous efficiency, presumably because human visual sense has been developed with a special reference to sunlight. Eddington took the maximum of $P$ as a standard and set the bolometric magnitude at this effective temperature to agree with the visual magnitude. 

At any other temperature, $P$ will be smaller, and the star will be brighter bolometrically than visually. The following table (Table \ref{tab:tab2}) is given by \citet{Eddington1926} to display effective temperatures and corresponding $P$ values, which were converted to the light deficit 
$\Delta m$ as $m_{\rm v}-m_{\rm Bol}$ in the last column using the formula:
\begin{equation}
\label{eq:03}
\Delta m = -\frac{5}{2} \log P. 
\end{equation}

\begin{table}[h!]
\renewcommand{\tabcolsep}{7.9mm}
\renewcommand{\arraystretch}{1.1}
\centering
\caption{Effective temperature, and corresponding luminous efficiency and reduction of bolometric to visual magnitude \citep[credit to][]{Eddington1926}.}
\begin{tabular}{|c|c|c|}
\hline
$T_{\rm e}$ & $P$    & $\Delta m$~(Vis - Bol)  \\ \hline  
~2 540  & 0.092  & 2.59  \\ 
~3 000  & 0.206  & 1.71  \\ 
~3 600  & 0.417  & 0.95  \\ 
~4 500  & 0.723  & 0.35  \\ 
~6 000  & 1.000  & 0.00  \\ 
~7 500  & 0.985  & 0.02  \\ 
~9 000  & 0.893  & 0.12  \\
10 000  & 0.749  & 0.31  \\
12 000  & 0.616  & 0.53  \\
\hline
\end{tabular}
\label{tab:tab2}
\end{table}

Eddington came very close to defining the bolometric correction of a star by adding the third column of Table~\ref{tab:tab2}, which indicates the magnitude difference of a star between its visual and bolometric brightness, which he named the light deficit $\Delta m$. Eddington added this column for those who cannot access bolometric measurements of a star, but only it's visual. Suppose a star has a visual brightness and a parallax. The mass could still be obtained from Figure~\ref{fig:01} or Table~\ref{tab:tab1} after its absolute visual magnitude converted to absolute bolometric magnitude through Table 2. The luminous efficiency (column 2) and the light deficit (column 3) in Table \ref{tab:tab2} also have the potential to indicate the stellar absolute temperature scale alone. Eddington did not notice it because his primary concern was to establish his famous mass-luminosity law.      

\subsection{The intermediate concept: Heat index}
Unfortunately, the concept of luminous efficiency did not continue after \citet{Eddington1926}. Instead, the concept of ``Heat-index'' was defined by \citet{Pettit1928} in a paper where radiometric and water cell absorption observations of 124 stars during the period 1922 to 1927 were discussed. First, radiometric magnitude is defined as equal to the apparent magnitude of an A0 star, which will give the observed galvanometer deflection. Next, ``Heat-index'' is defined as the quantity of the difference between visual and radiometric magnitude ($m_{\rm v}-m_{\rm r}$). The scale of bolometric magnitudes was made to coincide with the radiometric magnitudes at G0, which differed from Eddington’s scale by about 0.1 mag. The phrase ``A0 stars were chosen for the determination of the zero point since photographic and visual magnitudes are equal for this spectral class; and, since the heat-index is visual minus radiometric magnitude, the heat-index, like the colour index, is also zero for these stars.'' from \citet{Pettit1928} indicated that like the observed colour index of stars, their heat index could be used as stellar temperature indicators. 

\citet{Pettit1928} gave a table of 124 stars giving their spectral types, absolute visual magnitudes ($M_{\rm v}$), Harvard apparent visual ($m_{\rm{v}}$) and radiometric ($m_{\rm{r}}$) magnitudes, heat-index, water cell absorption, correction to no atmosphere ($\Delta m_{\rm{r}}$) including losses in the telescope, total radiation in units of $\rm{cal\ cm^{-2}\ min^{-1}}$ incident outside the atmosphere, absolute bolometric magnitude ($M_{\rm{B}}$), absolute temperature ($T$) from the heat-index and water cell absorption. It could be understood from this paper that the total radiation predicted from radiometric magnitudes of no atmosphere, bolometric magnitudes, and water cell absorption are mainly for setting up the stellar absolute temperature scale. The absolute bolometric magnitude $M_{\rm{B}}$ of a star was calculated by the formula,
\begin{equation}
\label{eq:04}
M_{\rm{B}} = m_{\rm{r}} - \Delta m_{\rm{r}} + 5 \log \varpi + 5.9 
= M_{\rm{V}} - H.I. - \Delta m_{\rm{r}} + 0.9, 
\end{equation}
so that the apparent bolometric magnitude of a star comes from its radiometric magnitude with atmospheric corrections, $m_{\rm{b}}= m_{\rm{r}}-\Delta m_{\rm{r}}+0.9$ if its parallax ($\varpi$) was known, thus, $M_{\rm{V}}$ is the absolute visual magnitude and H.I. is the heat-index. The constant (0.9) was used to fix the zero point of bolometric magnitudes to the giants in spectral classes F5 to G5. The term “bolometric correction” ({\it BC}) was not also used by \citet{Pettit1928}, who gave the radiometric magnitude of the Sun -27.18 mag, which indicates the heat-index of the Sun is 0.46 mag, if Russell’s value of the Sun’s visual magnitude, -26.72 was used.

To convert the visual absolute magnitude ($M_{\mathrm{v}}$) into bolometric values, \citet{Stromberg1932} used the same formula: $M_{\rm b} = M_{\rm v} - H.I. - \Delta m_{\rm r} + 0.9$ similarly. But, he suggested a new term $(M_{\rm{b}} - M_{\rm{v}})$ besides the heat-index and presented a table giving spectral types from B1 to dM2 including the Sun in the first column. The second column gave corresponding temperatures, and the other columns included corrections to no atmosphere, heat-index, and $(M_{\rm{b}}-M_{\rm{v}})$. Contemporary astrophysicists would immediately recognize the term $(M_{\rm{b}} - M_{\rm{v}})$ as bolometric correction. However, \citet{Stromberg1932}, like astrophysicists before him, did not prefer to use the term ``bolometric correction'' but instead referred to it as ``the resulting corrections, $M_{\rm{b}} - M_{\rm{v}}$'' when describing his table. He also gave the formula:
\begin{equation}
\label{eq:05}
\log E = \log E_{\rm{s}} - 0.4 \, (M_{\rm{b}} - M_{\rm{s}}), 
\end{equation}
where $M_{\rm{s}} = 4.84$ is the bolometric absolute magnitude of the Sun, $M_{\rm b}$ is that of the star, and $E_{\rm{s}}$ is the total power output of the Sun, $E_{\rm{s}} = 3.79 \times 10^{33}$ ergs s$^{-1}$.

\subsection*{The Final Concept: Bolometric Correction} 
The term ``bolometric correction'' was first used by \citet{Kuiper1938a}, who claimed that there must be a single effective temperature defined by the Stefan-Boltzmann law as:
\begin{equation}
\label{eq:06}
L = 4 \pi R^2 \cdot \sigma T_{\rm e}^4, 
\end{equation}
in which $L$ is the luminosity, $R$ is the radius, and $\sigma$ is the radiation constant. Thus, $\sigma T_{\rm e}^4$ is the total surface flux (at all wavelengths) of a black body with a unique effective temperature $T_{\rm e}$ for a star. Apparently, \citet{Kuiper1938a} began his discussions by correcting \citet{Eddington1926}, who had claimed two effective temperatures: $T_{\rm e}$ from the quantity of the radiant energy according to the Planck law as the first and $T_{\rm e}'$ from Wien’s law as the second. Kuiper’s suggestion was commonly accepted, and even today. Equation~(\ref{eq:06}) was started to be used as the standard definition of the effective temperature, and since then, the temperature $T_{\rm e}'$ from Wien’s law is recognized as the colour temperature.
The primary concern of Kuiper, therefore, appears to be the determination of effective temperatures of stars. As the common purpose of astronomers of that time, he said: ``If $L$ and $R$ are known, $T_{\rm e}$ may be derived from Equation~(\ref{eq:06}). But since $L$ is never known directly from observation, and usually only the absolute visual (or photovisual) magnitude is available, Equation~(\ref{eq:06}) can be used in practice only if a table of bolometric corrections is available.'' Consequently, Kuiper defined the bolometric correction of a star as:
\begin{equation}
\label{eq:07}
{\rm B.C.} = M_{\rm bol} - M_{\rm pv} = m_{\rm bol} - m_{\rm pv}, 
\end{equation}

if
\begin{equation}
\label{eq:08}
M_{\rm bol} = M_{\rm bol,\odot} - 2.5 \log L. 
\end{equation}

The Equation~(\ref{eq:08}) is valid only if $L$ is in solar units. That is, the equation could also be written as: $M_{\rm bol} = M_{\rm bol, \odot} - 2.5 \log \left(\frac{L}{L_\odot} \right)$, where $L$ and $L_\odot$ could be in any units. This formula is not different from Equation~(\ref{eq:05}) above \citep{Stromberg1932}, where $L$ was denoted by $E$.

Here, \citet{Kuiper1938a} was saying bolometric corrections of stars are independent of their distances, thus, $M_{\rm bol}-{M_{\rm pv}}= m_{\rm bol}-m_{\rm pv}$. If a {\it BC} is available for a star, its $M_{\rm bol}$ can be deduced by adding the {\it BC} to its $M_{\rm pv}$. Once $M_{\rm bol}$ is ready (possible only if the parallax of the star were measured even if it was from radiometric observations), then the $L$ of the star in solar units could be computed by Equation~(\ref{eq:08}). If the $R$ of the star is also known, which could be from interferometric observations or deduced from eclipsing light curves, Equation~(\ref{eq:06}) could be used for deducing the effective temperatures of the stars. Thus, ``A satisfactory table of bolometric corrections can be constructed without recourse to the assumption of black-body radiation (even without any need of radiometric measurements), and that also the scale of stellar effective temperatures can be brought into a fairly satisfactory state'' said by \citet{Kuiper1938a} to confirm his first sentence in the article ``The purpose of this article is the selection or derivation of the most probable set of effective temperatures and bolometric corrections that can now be obtained. These data are required in the discussion of mass-luminosity relation given in the following paper, '' which refers to \citet{Kuiper1938b}.  

By revising and examining all radiometric, photographic, and visual observations of stars and the Sun, including those of \citet{Pettit1928}, he ended up a conclusion with: $m_{\rm bol,\odot} = -26.95, \quad M_{\rm bol,\odot} = 4.62, \quad m_{\rm pv,\odot} = -26.84, \quad M_{\rm pv,\odot} = 4.73 $, mags, indicating $BC_\odot = -0.11$ mag. Those basic data were then used by him to establish his table of effective temperatures and bolometric corrections for stars with effective temperatures ranging from 3000 to 50 000 K, as given in Table \ref{tab:tab3}. To avoid bolometric corrections with different signs, $0.10$ mag was subtracted from all values. Thus, all {\it BC} values in \citet{Kuiper1938a} are negative in Table \ref{tab:tab3}. The apparent photovisual magnitude of the Sun was later revised by \citet{Stebbins1957}, who introduced six-colour photometry ($U$, $B$, $V$, $G$, $R$, $I$) and determined: $m_{\rm pv,\odot}=-27.73 \pm 0.03$ mag by comparing solar light to stars and a standard lamp.

\begin{table}[h!]
\renewcommand{\tabcolsep}{7.5mm}
\renewcommand{\arraystretch}{1.1}
\centering
\caption{Stellar effective temperatures and bolometric corrections by \citet{Kuiper1938a}.}
\vspace*{-2pt}
\begin{tabular}{lccc}
\multicolumn{4}{c}{Theoretical Values of Bolometric Corrections}\\
\multicolumn{4}{c}{(Temperatures in 1000$^{\rm o}$)}\\
\hline
$T_{\rm e}$ & {\it BC} & $T_{\rm e}$& {\it BC}      \\ 
\hline  
3.0  & -$3^{\rm m}.2$  & 13        & -$1^{\rm m}.18$ \\
3.5  & -2.10           & 14        & -1.35          \\
4.0  & -1.30           & 15        & -1.51          \\
4.5  & -0.63           & 16        & -1.66          \\
5.0  & -0.34           & 17        & -1.80          \\
5.5  & -0.17           & 18        & -1.94          \\
6.0  & -0.06           & 19        & -2.06          \\
6.5  & -0.00           & 20        & -2.18          \\
7.0  & -0.01           & 22        & -2.40          \\
7.5  & -0.12           & 25        & -2.69          \\
8.0  & -0.22           & 30        & -3.12          \\
9.0  & -0.40           & 35        & -3.50          \\
10   & -0.57           & 40        & -3.80          \\
11   & -0.78           & 45        & -4.10          \\
12   & -0.98           & 50        & -4.30         \\
\hline       
\end{tabular}
\label{tab:tab3}
\end{table}

A similar table (Table \ref{tab:tab4}) indicating the primary purpose of calculating bolometric corrections is to set up a temperature scale for the spectral types of stars, where three different bolometric corrections were compared, is also prepared by \citet{Lohmann1948}. Notice again that all {\it BC} values are negative. 

\begin{table*}[h!]
\renewcommand{\tabcolsep}{3mm}
\renewcommand{\arraystretch}{1.1}
\centering
\caption{Spectral types, effective temperatures, and bolometric corrections \citep[credit to][]{Lohmann1948}.}
\begin{tabular}{cccccc}
\multicolumn{6}{c}{Die bolometrische Korrektion} \\
 \toprule
 Spektrum & Farbtemperatur & $M_{\rm {vis}}$ & \multicolumn{3}{c}{bolometrische Korrektion} \\
 \cmidrule(lr){4-6}
 & & & LOHMANN & KUIPER & EDDINGTON-PIKE      \\
 \midrule
 B0  &  33 000$^{\rm o}$ &     & -3.06 & -2.70 & -2.9 \\
 B5  &  24 000 &       & -2.00 & -1.58 & -2.0 \\
 A0  &  16 000 &       & -0.98 & -0.72 & -1.2 \\
 A5  &  11 000 &       & -0.37 & -0.31 & -0.4 \\
 F0  & ~~8 400 & +1.7  & -0.04 & -0.00 &  0.0 \\
 F5  & ~~6 900 & +1.0  & ~0.00 & -0.04 &  0.0 \\
 G0  & ~~6 100 & -0.2  & ~0.00 & -0.25 &  0.0 \\
 G5  & ~~5 500 & +0.3  & -0.10 & -0.39 &  0.0 \\
 K0  & ~~4 800 & ~~0.0 & -0.28 & -0.54 & -0.2 \\
 K5  & ~~4 100 & ~~0.0 & -0.78 & -1.35 & -0.6 \\
 M0  & ~~3 600 & -0.3  & -1.25 & -1.55 & -0.9 \\
 M5  & ~~3 100 & -0.5  & -1.90 & -3.40 & -1.6 \\
 \bottomrule

\end{tabular}
\label{tab:tab4}
\end{table*}

\subsection{The First Critic; Similarities, and Dissimilarities of the Three Suggestions}

From a mathematical point of view, all three suggestions above are the same because they all indicate a difference between the visual and the bolometric magnitudes of a star. The same entity was called ``luminous efficiency'' for the first, ``heat-index'' for the second to indicate visual-bolometric, while “bolometric correction” for the third to imply bolometric-visual, which may appear different, but in an absolute sense it is the same. However, the literal names given to the same entity recall different concepts physically. 

Luminous efficiency appears to be an explanation for why human eyes (or one of the filters in a photometric system in today’s terms) see stars cooler or hotter than the Sun dimmer in the visual than the bolometric despite having the same luminosity. The heat index, on the other hand, emphasises bolometric magnitudes coming from the heat measurements of the stars, which could be measured by a galvanometer, bolometer, or radiometer. In another saying, the term ``heat index'' reminds a reader that bolometric magnitudes are observable quantities. 

Contrarily, the term ``bolometric correction'' implies that the bolometric magnitude of a star is an abstract quantity, unobservable like the luminosity of a star [remember \citet{Kuiper1938a}'s comment:``... since $L$ is never known directly from observation, but usually only the absolute visual (or photovisual) magnitude... '' above]. An eye can detect visual photons. Thus, visual magnitudes are measurable, while photographic plates could also detect additional photons other than the eye can see, thus, photographic magnitudes are also measurable. Because there is no telescope or a detector to detect photons of all frequencies, representing the total radiation of a star, its bolometric magnitude could also be unobservable, like its $L$. Does this mean \citet{Kuiper1938a}'s definition of {\it BC} rejects using radiometric measurements for determining {\it BC} of a star? The answer is no. ``... we denote with $\Delta m$, bolometric corrections computed on the assumption of black-body radiation” and “… It is well known that the stars do nor radiate as black bodies. Therefore, the values $\Delta m$ can not be used; they need a correction, which we denote by $\delta m$. This correction can be evaluated only if the whole spectral energy curve is known. Two approximations are available to this ideal case: (I) theoretical spectral energy curves, computed based on a certain assumed composition of stellar atmospheres, and (II) radiometric observations of low-temperature stars (for high temperatures, too much energy is radiated in the inaccessible ultraviolet” are two critical comments from \citet{Kuiper1938a} for us to understand the details how and why he defined and produced bolometric corrections. 

According to \citet{Pettit1928}, $BC=m_{\rm b}-V=(m_{\rm r} -\Delta m_{\rm r}+K)-V$ where $m_{\rm b}$ and $V$ are apparent bolometric and visual magnitudes, $m_{\rm r}$ and $\Delta m_{\rm r}$ are apparent radiometric magnitude and its correction to no atmosphere, including losses in the telescope as explained and cited by \citet{Eggen1956}, who adopted a value of 0.62 for the constant $K$, which establishes the zero-point of the {\it BC} in the magnitude scale.   

Nevertheless, the word ``correction'' implies a missing part or a deficit. Since $BC=M_{\rm bol}-M_{\rm V}= m_{\rm bol}-V$, then, $M_{\rm bol}= M_{\rm V} + BC$ and/or $m_{\rm bol}= V + BC$, indicates visual magnitudes have a deficit, and if this deficit  is added to $V$, the total brightness (a brightness without any deficit), $M_{\rm bol}$ and/or $m_{\rm bol}$ of the star representing photons of at all wavelengths from zero to infinity could be obtained. If this is the case, one may think that if visual magnitudes have deficits, why not photographic magnitudes? According to this argument, the bolometric magnitude of a star could also be obtained from its photographic magnitude as $M_{\rm bol}= M_{\rm P} + BC_{\rm p}$ or $m_{\rm bol}= m_{\rm p} +BC_{\rm p}$ if the deficit of the photographic magnitudes ($BC_{\rm p}$) were known. We did not see \citet{Kuiper1938a} to define $BC_{\rm p}$, he only defined $BC$ for the visual. 

A possible reason was that the photographic magnitudes were assumed monochromatic because they were measured from the sizes of star images on the black and white photographs of that time. Indeed \citet{Kuiper1938a} commented, ``It is seen that these corrections to photovisual magnitudes are very similar to those for visual magnitudes, although photovisual magnitudes are more nearly monochromatic''. Photovisual magnitudes being similar to visual magnitudes must mean the opposite, that is, photovisual magnitudes cannot be considered monochromatic despite coming from black-and-white images, which is also true for photographic magnitudes. Defining {\it BC} only for visual, and the above comment of \citet{Kuiper1938a}, imply that a photographic magnitude was considered monochromatic, and thus, it was meaningless for him to discuss a deficit for it.

Naming the same mathematical quantity (the difference between visual and bolometric mags) with dissimilar concepts has special implications for the zero-point problem of bolometric corrections and for the other problems that had become misleading paradigms by then.

\subsection{Further Developments}
A new formulation on the line of development for computing {\it BC} from a model atmosphere is given by \citet{McDonald1952} as
\begin{equation}
    B.C.=2.5\log \frac{\int P_{\lambda}F_{\lambda}d_{\lambda}}{\int P_{\lambda}F_{\lambda}({\odot})d_{\lambda}}-10\log\frac{T_{\rm e}}{T_{\rm e}(\odot)}-0.11,
    \label{eq:09}
\end{equation}
where the bolometric correction of a star is to a scale giving $BC=-0.11$ for the Sun in which $P_{\lambda}$ is the normalised visual sensitivity function and $F_{\lambda}$ is the surface monochromatic flux. It appears new and different, but it is not. It is the combination of Equations~(\ref{eq:02}) and (\ref{eq:03}); that is, it is the logarithmic form of $P$, the luminous efficiency, as it was called before. The only difference here is that it was expressed relative to the Sun’s luminous efficiency, and the quantity $\int I^{'}(\lambda, T_{\rm e})d\lambda$ (the total radiation output of the star) is replaced by $\sigma T^{4}_{\rm e}$ (bolometric flux on the surface of the star). The symbol $\sigma$ cancels out in the division if $P$ is written for the star and the Sun. Two things appear not right here: 1) the quantity $\int I^{'}(\lambda, T_{\rm e})d\lambda$ should have been replaced by $L=4\pi R^2\cdot \sigma T^{4}_{\rm e}$ not only $\sigma T^{4}_{\rm e}$ because there could be size difference between the star and the Sun. 2) {\it BC} should not be a relative quantity because it is impractical to use it for a star as a relative quantity.

\begin{figure}
\centering
\includegraphics[width=0.98\linewidth]{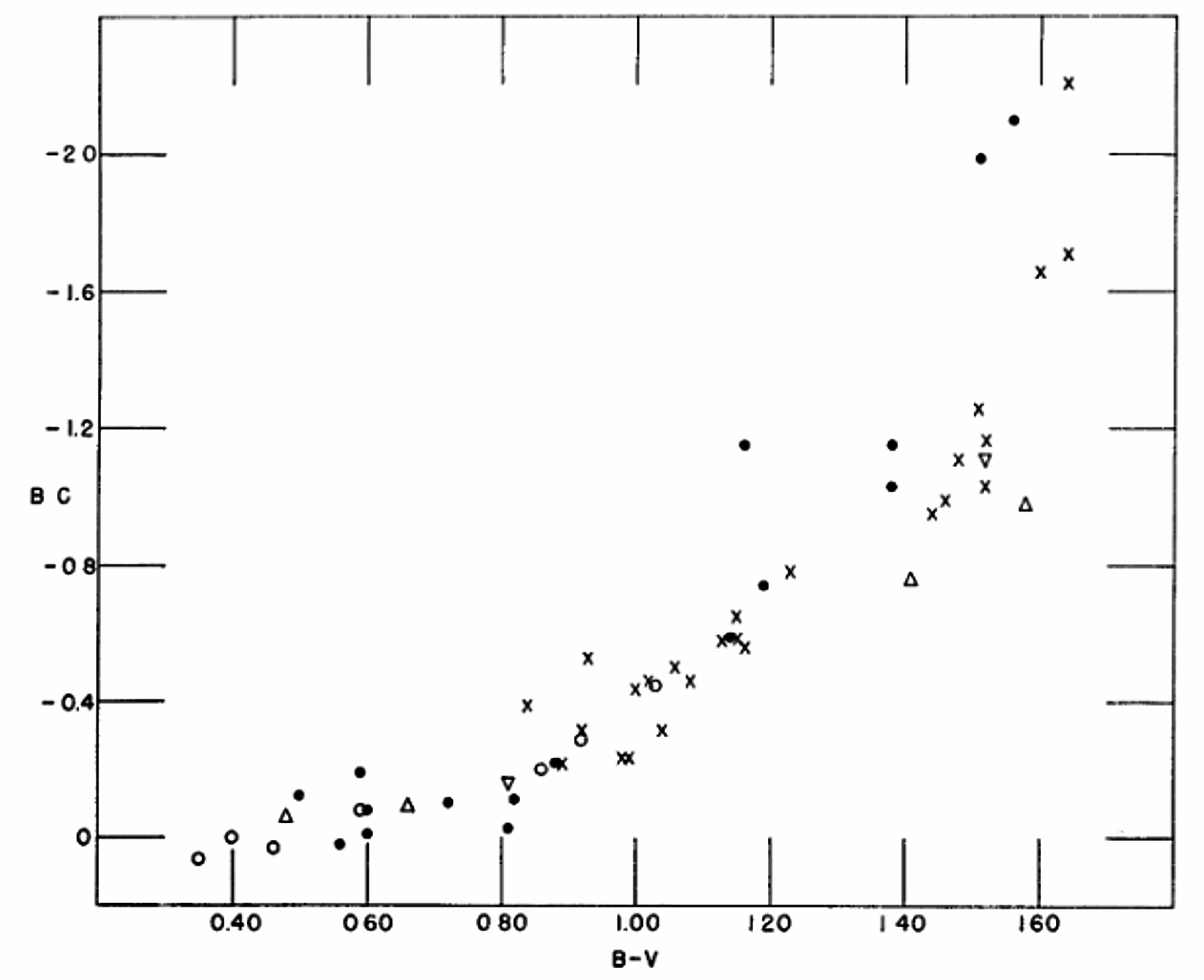}
\caption{{\it BC} from radiometric and photoelectric magnitudes. Dots: luminosity class V; circles: IV; crosses: III; $\nabla$: II; $\delta$: I \citep[credit to][]{Popper1959}.}
\label{fig:02}
\end {figure}

The next most comprehensive study of bolometric corrections after \citet{Kuiper1938a} is \citet{Popper1959} which changed the tradition of giving {\it BC} against $T_{\rm eff}$. Popper preferred to display {\it BC} against $B-V$ rather than $T_{\rm eff}$. Figure~\ref{fig:02} displays {\it BC} from radiometric and photoelectric magnitudes from \citet{Popper1959}.  

\begin{figure*}
\centering
\includegraphics[width=0.98\linewidth]{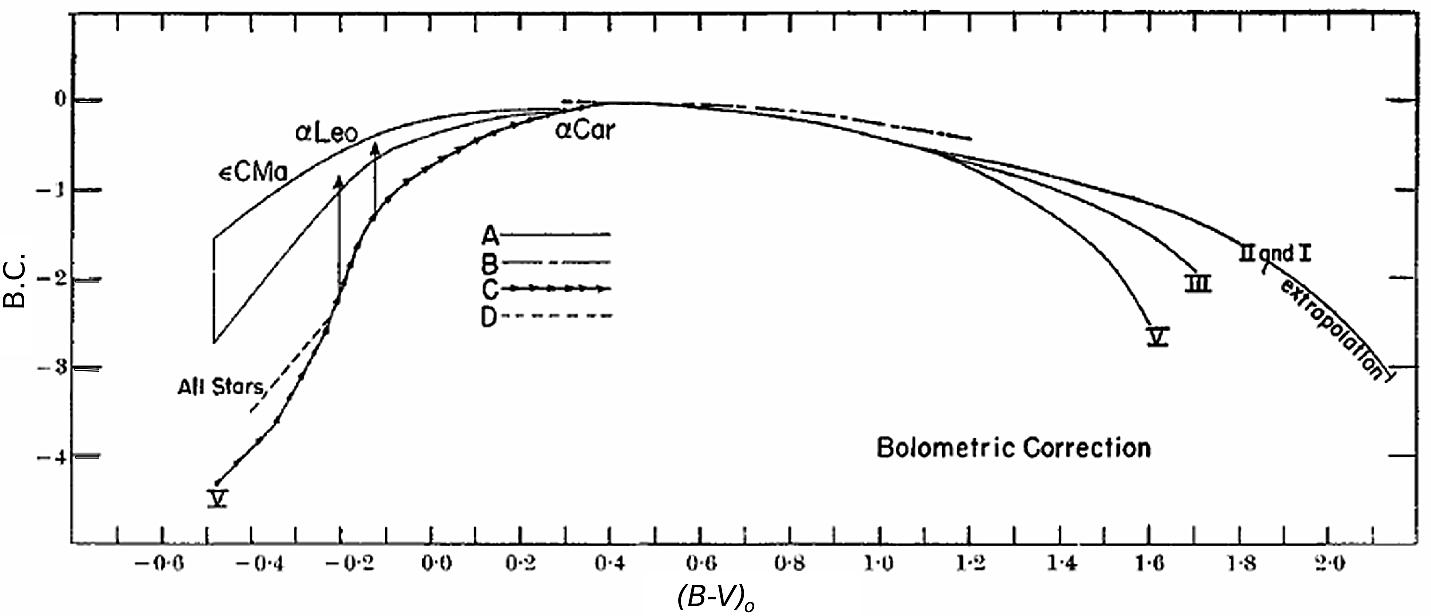}
\caption{Bolometric correction scales during 1960's are compared on {\it BC}-$(B-V)_{\rm o}$ plane \citep{Wildey1963}.} 
\label{fig:03}
\end {figure*}

``The zero point of the bolometric corrections is chosen so that Kuiper’s values for main-sequence stars of spectral types G0-G8 are produced.'' ``With the choice of zero point, the bolometric corrections of for individual stars are given by $BC=m_{\rm r}-\Delta m_{\rm r}-V+0.58$, where $m_{\rm r}$ and $\Delta m_{\rm r}$ are as defined by \citet{Pettit1928}. The $\Delta m_{\rm r}$ cannot be evaluated for stars hotter than about type F0 because of the unknown ultraviolet fluxes'', ``Bolometric correction for a star of the same colour as the Sun, $B-V=0.63$, is -0.07 mag'' and ``For the hotter stars, where {\it BC} have not yet been observed because of absorption in the earth’s atmosphere, one must compute {\it BC} from stellar atmospheric modes'' are some of the critical quotes from \citet{Popper1959}. Various {\it BC} scales which all without any positive {\it BC} are compared on $BC-(B-V)_{\rm 0}$ plane,  including the ones from Popper are displayed by \citet{Wildey1963} in Figure \ref{fig:03}.  

With the infrared photometric program, whose first report was published by \citet{Johnson1962}, the observing window of astronomers increased from 0.35 microns in the ultraviolet to 10 microns in the infrared. Subsequently, astronomers began using filters representing 10 different wavelength bands, which was defined by Johnson himself as $UBVRIJKLMN$. Computed {\it BC}s are given in separate tables containing spectral types as the first column and followed by $U-B$, $B-V$, $V-R$, $V-I$, $V-J$, $V-K$, $V-L$, $V-M$, and $V-N$ intrinsic colours, {\it BC} values and effective temperatures. The early results from the bright stars are given by \citet{Johnson1964}. Then, finalized results with additional observations at infrared are published by \citet{Johnson1966}.

According to \citet{Johnson1964, Johnson1966}, the bolometric correction is defined as $m_{\rm b}-V$, where $m_{\rm b}$ and $V$ are apparent bolometric and visual magnitudes of a star, and it is the correction to be applied to $V$ to obtain the apparent bolometric magnitude. The apparent bolometric magnitude is basically integrated absolute intensity expressed in mag units. The zero points were set to make $BC_{\odot}=0.0$ for the Sun. Bolometric corrections for A-type stars were found to be smaller (more positive) than those obtained before \citep{Kuiper1938a, Popper1959} for such stars. These new values of {\it BC} are from rocket observations at 1314\AA~and 1427\AA~by \citet{Chubb1963}. For the stars cooler than the Sun, the values of {\it BC} were computed and are listed in the next-to-last columns of the three tables containing giants, main sequence, and supergiants, respectively. Only the table with main-sequence stars is shown here (Table~\ref{tab:tab5}).  From the presented papers and the comment ``... all of the bolometric corrections listed in the tables of this chapter are based upon actual observational data'' it is not possible for one to understand whether or not Johnson used radiometric measurements or just only photometric data. Nevertheless, \citet{Johnson1964, Johnson1966} broke the old tradition and published tables containing both positive and negative {\it BC} values (see Table \ref{tab:tab5} here).

\begin{table*}[h]
    \renewcommand{\tabcolsep}{2.5mm}
    \renewcommand{\arraystretch}{1.0}
    \centering
    \small
     \caption{Spectral types, intrinsic colours, bolometric corrections and effective temperatures of main-sequence stars \citep[credit to][]{Johnson1966}.}
    \begin{tabular}{l c c c c c c c c c c c c}
        \toprule
        Sp(V) & $U-V$ & $B-V$ & $V-R$ & $V-I$ & $V-J$ & $V-K$ & $V-L$ & $V-M$ & $V-N$ & {\it B.C.} & $T_{\rm e}$~($^{\rm o}$K) \\
        \midrule
O5-7  & -1.46 & -0.32 & -0.15 & -0.47 & -0.73 & -0.94 & -1.01  & ---   & ---   & ---   & ---    \\
O8-9  & -1.44 & -0.31 & -0.15 & -0.47 & -0.73 & -0.94 & -1.01  & ---   & ---   & ---   & ---     \\
O9.5  & -1.40 & -0.30 & -0.14 & -0.46 & -0.73 & -0.94 & -1.00  & ---   & ---   & ---   & --- \\
B0    & -1.38 & -0.30 & -0.13 & -0.42 & -0.70 & -0.93 & -0.99  & ---   & ---   &-2.27  & 26500 \\
B0.5  & -1.29 & -0.28 & -0.12 & -0.39 & -0.66 & -0.88 & -0.93  & ---   & ---   &-1.92  & 24500 \\
B1    & -1.19 & -0.26 & -0.11 & -0.36 & -0.61 & -0.81 & -0.86  & ---   & ---   &-1.59  & 21500 \\
B2    & -1.10 & -0.24 & -0.10 & -0.32 & -0.55 & -0.74 & -0.77  & ---   & ---   &-1.31  & 18000 \\
B3    & -0.91 & -0.20 & -0.08 & -0.27 & -0.45 & -0.61 & -0.63  & ---   & ---   &-1.00  & 13500 \\
B5    & -0.72 & -0.16 & -0.06 & -0.22 & -0.35 & -0.47 & -0.48  & ---   & ---   &-0.70  & 13800 \\
B6    & -0.63 & -0.14 & -0.06 & -0.19 & -0.30 & -0.41 & -0.41  & ---   & ---   &-0.56  & 12900 \\
B7    & -0.54 & -0.12 & -0.04 & -0.17 & -0.25 & -0.35 & -0.34  & ---   & ---   &-0.48  & 12200 \\
B8    & -0.39 & -0.09 & -0.02 & -0.12 & -0.17 & -0.24 & -0.22  & ---   & ---   &-0.30  & 11300 \\
B9    & -0.25 & -0.06 &  0.00 & -0.06 & -0.09 & -0.14 & -0.11  & ---   & ---   &-0.14  & 10600 \\
A0    &  0.00  & 0.00  &+0.02  & 0.00 & -0.01 & -0.03 &  0.00  & -0.03 & -0.03 & -0.01 & ~~9850 \\
A2    & +0.12 & +0.06 & +0.08 & +0.09 & +0.11 & +0.13 & +0.16  & +0.13 & +0.13 & +0.08 & ~~9120 \\
A5    & +0.25 & +0.14 & +0.16 & +0.22 & +0.27 & +0.36 & +0.40  & +0.36 & +0.36 & +0.13 & ~~8260 \\
A7    & +0.30 & +0.19 & +0.19 & +0.28 & +0.35 & +0.46 & +0.52  & +0.46 &+0.46  & +0.13 & ~~7880 \\
F0    & +0.37 & +0.31 & +0.30 & +0.47 & +0.58 & +0.79 & +0.86  & +0.79 &+0.79  & +0.11 & ~~7030 \\
F2    & +0.39 & +0.36 & +0.35 & +0.55 & +0.68 & +0.93 & +1.07  & +0.93 &+0.93  & +0.09 & ~~6700 \\
F5    & +0.43 & +0.43 & +0.40 & +0.64 & +0.79 & +1.07 & +1.25  & +1.07 &+1.07  & +0.06 & ~~6400 \\
F8    & +0.60 & +0.54 & +0.47 & +0.76 & +0.96 & +1.27 & +1.45  & +1.27 &+1.27  & +0.04 & ~~6000 \\
G0    & +0.70 & +0.59 & +0.50 & +0.81 & +1.03 & +1.35 & +1.53  & +1.35 &+1.35  & 0.00  & ~~5900 \\
G2    & +0.79 & +0.63 & +0.53 & +0.86 & +1.10 & +1.44 & +1.61  & +1.44 &+1.44  & 0.00  & ~~5770 \\
G5    & +0.86 & +0.66 & +0.54 & +0.89 & +1.14 & +1.49 & +1.67  & ---   & ---   & -0.02 & ~~5660 \\
G8    & +1.06 & +0.74 & +0.58 & +0.96 & +1.24 & +1.63 & +1.85  & ---   & ---   & -0.06 & ~~5440 \\
K0    & +1.29 & +0.82 & +0.64 & +1.06 & +1.38 & +1.83 & +2.00  & ---   & ---   & -0.12 & ~~5240 \\
K2    & +1.60 & +0.92 & +0.74 & +1.22 & +1.57 & +2.15 & +2.24  & ---   & ---   & -0.23 & ~~4960 \\
K5    & +2.18 & +1.15 & +0.99 & +1.62 & +2.04 & +2.75 & +2.84  & ---   & ---   & -0.55 & ~~4400 \\
K7    & +2.52 & +1.30 & +1.15 & +1.93 & +2.36 & +3.21 & +3.40  & ---   & ---   & -0.82 & ~~4000 \\
M0    & +2.67 & +1.41 & +1.28 & +2.19 & +2.71 & +3.60 & +3.78  & ---   & ---   & -1.10 & ~~3750 \\
M1    & +2.70 & +1.48 & +1.40 & +2.45 & +3.06 & +3.95 & +4.15  & ---   & ---   & -1.38 & ~~3600 \\
M2    & +2.69 & +1.52 & +1.50 & +2.69 & +3.37 & +4.27 & +4.47  & ---   & ---   & -1.64 & ~~3400 \\
M3    & +2.70 & +1.55 & +1.60 & +2.94 & +3.66 & +4.57 & +4.79  & ---   & ---   & -1.85 & ~~3300 \\
M4    & +2.70 & +1.56 & +1.70 & +3.19 & +3.97 & +4.87 & +5.20  & ---   & ---   & -2.17 & ~~3200 \\
M5    & +2.80 & +1.61 & +1.80 & +3.47 & +4.28 & +5.17 & (+5.54)& ---   & ---   & -2.48 & ~~3100 \\
M6    & +2.99 & +1.72 & +1.93 & +3.96 & +4.63 & +5.58 & (+6.03)& ---   & ---   & -2.82 & ~~2950 \\
M7    & +3.24 &  +1.84 &+2.20 & +4.20 & +5.20 & +6.18 &   ---  & ---   & ---   & -2.35 & (2850) \\
M8    &(+3.50)& (+2.00)&(+2.50)&(4 .70)&(+5.80) &(+6.75)& ---   & ---   &---   &  -3.9 & (2750) \\
     \hline
    \end{tabular}
    \label{tab:tab5}
\end{table*}

Publications discussing {\it BC} and effective temperature scales continued \citep{Lee1970, Heintze1973, Code1976, Flower1977, Hayes1978, Habets1981, Malagnini1985, Fitzpatrick1990, Napiwotzk1993} with an increasing rate approximately up to the time of \citet{Flower1996}, who gave intrinsic $(B-V)_{\rm 0}$ colour and empirical {\it BC} coefficients as functions of effective temperatures first time in the line of development. \citet{Flower1996} used 335 stars to calibrate {\it BC}, $(B-V)_{\rm 0}$ and $T_{\rm eff}$ scales, among them 297 with measured effective temperatures and 122 with reliable unreddened $B-V$ colours. \citet{Flower1996} gave two analytical $(B-V)_{\rm 0}-T_{\rm eff}$ relations $[(B-V)_{\rm 0}=a+b\log T_{\rm eff}+c(\log T_{\rm eff})^2+...]$, one for main sequence, giants and subgiants, and one for supergiants. Then, he gave three $BC-T_{\rm eff}$ relations $[BC=a+b\log T_{\rm eff}+c(\log T_{\rm eff} )^2+...]$ each valid for three temperature ranges ($\log T_{\rm eff}\geq3.9$ K, $3.9<\log T_{\rm eff}~{\rm (K)}<3.7$, $\log T_{\rm eff}~{\rm (K)}\leq3.7$ K), respectively, but usable for all luminosity classes.

\citet{Cayrel1997} investigated the behavior of the computed {\it BC} of the {\it Hipparcos} $H_{\rm p}$ band ($BC_{\rm Hp} = M_{\rm bol}- H_{\rm p}$) with basic physical parameters $T_{\rm eff}$, $\log g$ and, [Fe/H] using model atmosphere fluxes computed with Kurucz’s ATLAS9 code. \citet{Bessell1998} have computed broadband colours, bolometric corrections and temperature calibrations for all spectral types (O to M) in the Johnson-Cousins-Glass system from the synthetic spectra from new model atmospheres available for the time (see references therein). \citet{Masana2006} have presented a method to determine effective temperatures, angular semi-diameters, and bolometric corrections for population I and II FGK type stars based on $V$ and 2MASS ($JHK_{\rm s}$) photometric system \citep{Skrutskie2006} for the temperature range 4000-8000 K. From the application to a large sample of FGK {\it Hipparcos} dwarfs and subdwarfs, \citet{Masana2006} provided calibrations for both effective temperature and bolometric correction as a function of $(V-K)_{\rm 0}$, [m/H] and $\log g$. Extensive tables of bolometric corrections and interstellar extinction coefficients for the WFPC2 and ACS (both WFC and HRC) photometric systems were derived from synthetic photometry covering a large range of effective temperatures, surface gravity, and metallicity by \citet{Girardi2008}. Intrinsic colours, temperatures and bolometric corrections of pre-main-sequence stars are discussed by \citet{Pecaut2013}. Empirical colour-colour relations, zero-age main sequence relations, Spt-$M_{\rm V}$, Spt-$T_{\rm eff}$, colour-Spt relations, and $T_{\rm eff}-BC$ relations derived from Sejong Open Cluster Survey (SOS) are presented as tables by \citet{Sung2013}. A database in which existing popular spectral libraries assembled to compute {\it BC} tables homogeneously for a large variety of photometric systems was given by \citet{Chen2019}. 

\citet{Flower1996}’s relations had serious printing errors as pointed out by \citet{Torres2010}. However, one frequently used source of {\it BC} is the {\it BC}-$T_{\rm eff}$ relation by \citet{Flower1996}. \citet{Torres2010} rectified the coefficients in the relation and republished them with a large number of digits (15). Moreover, he also corrected the misprint of the $(B-V)_{\rm 0}-T_{\rm eff}$ relations as $T_{\rm eff}-(B-V)_{\rm 0} [\log T_{\rm eff}=a+b(B-V)+c(B-V)^2 ...]$ and presented all coefficients in higher precision. Nevertheless, using 15-digit coefficients is really not practical for some users and rounding them to a smaller number of digits may cause a noticeable systematic shift to the {\it BC} values. It must be because of such complexities; some users may prefer to have tabulated {\it BC} tables. According to \citet{Lester2011}, the variation of {\it BC} with spectral types found by \citet{Johnson1964} is similar to the more modern curves therefore, in some ways, the photometric data of the bulletin paper is still fresh. 

\section{How and Why Misleading Paradigms Were Developed?}

\begin{figure}
\centering
\includegraphics[width=0.98\linewidth]{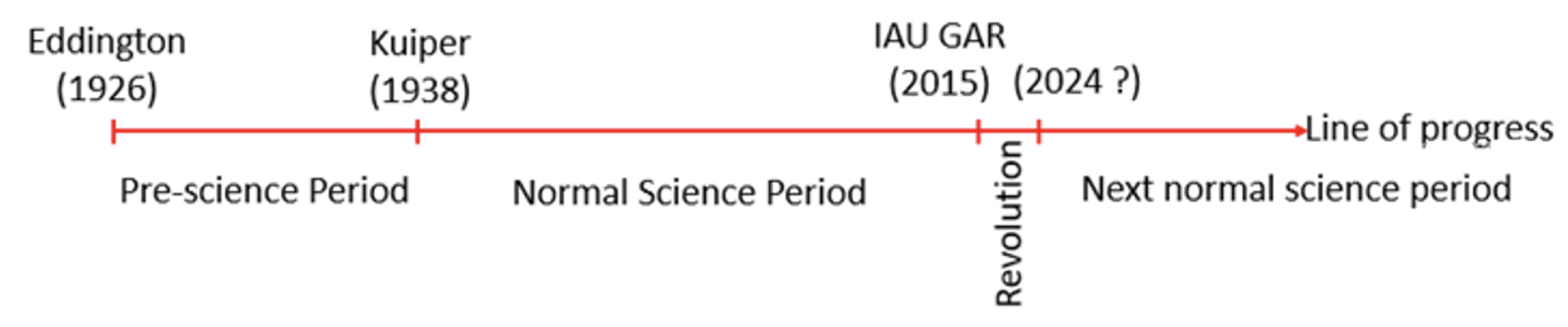}
\caption{Progress line of bolometric corrections according to Kuhnian philosophy of science.} 
\label{fig:04}
\end{figure}

Examining all of the tabulated tables of {\it BC} and the empirical $BC-T_{\rm eff}$ relations from \citet{Flower1996} and \citet{Eker2020}, it is possible to classify {\it BC} producers into two groups: 1) Ones who publish tabulated {\it BC} values without any positive value \citep{Kuiper1938a, McDonald1952, Popper1959, Wildey1963, Hayes1978, Habets1981, Cox2000, Pecaut2013} and 2) ones who published tabulated {\it BC} values, which are mostly negative but there could be a limited number of positive values 
\citep{Code1976, Johnson1964, Johnson1966, Flower1977, Flower1996, Bessell1998, Sung2013, Casagrande2018a, Casagrande2018b, Eker2020}.
It is possible to say the two groups appear as if the first group is obeying and the second group is disobeying the paradigms: 1) ``Bolometric corrections must always be negative'', 2)``the bolometric magnitude of a star ought to be brighter than its $V$ magnitude'' and 3) ``the zero point of bolometric corrections are arbitrary''. This is because it is quite possible that the first group is formed by {\it BC} producers who convinced themselves that {\it BC} of a star must always be negative as in the case of \citet{Kuiper1938a} who subtracted 0.10 mag from each of the computed {\it BC} values in order to avoid {\it BC} numbers with different signs. In another saying, the arbitrariness attributed to the zero-point constant of the {\it BC} scale gives the right to the first group to make sure their {\it BC} are all negative by systematically shifting their original {\it BC} by any constant number.  Nevertheless, a negative {\it BC} indicates both $M_{\rm bol} < M_{\rm V}$ and $m_{\rm bol} < V$, which verbally means that ``a star is brighter bolometrically than visually'' according to the definition $BC= M_{\rm bol}-M_{\rm V}=m_{\rm bol}-V$, first given by \citet{Kuiper1938a}.   
At this point, what appears problematic is the behaviour of the second group, which appears to be ignoring or rebelling against the paradigms. We will come to this problem later together with the problems associated with the paradigms listed above, which in reality appear to be forecasting ``the end'' of the first normal science period and/or the upcoming of the first revolution in the development line of the bolometric corrections (Figure~\ref{fig:04}), where the period before \citet{Kuiper1938a} was shown as the pre-science period.     

According to the Kuhn's philosophy of science, in the beginning, there could be numerous theories, hypotheses or worldviews which compete with each other. Thus, the period until one of them overcomes or dominates the others is called pre-science period. Assuming that each theory or hypothesis has its own paradigms, the paradigms associated with the victorious worldview become the paradigms of the first normal science period. Studying the pre-science period is important to understand how and why misleading paradigms were initiated and customized.         

\subsection{Initiations and Customizations of the First and the Second Paradigms}

\citet{Heintze1973} was the first who formulated {\it BC} of a star as
\begin{equation}\label{eq:10}
    \begin{aligned}
        BC &= 2.5 \log \left( \frac{\int_{0}^{\infty} S_{\lambda} \mathcal{F}_{\lambda} \, d\lambda}{\int_{0}^{\infty} \mathcal{F}_{\lambda} \, d\lambda} \right) + c = 2.5 \log \left( \frac{\int_{0}^{\infty} S_{\lambda} f_{\lambda} \, d\lambda}{\int_{0}^{\infty} f_{\lambda} \, d\lambda} \right) + c = 2.5 \log \left( \frac{\int_{0}^{\infty} S_{\lambda} F_{\lambda} \, d\lambda}{\int_{0}^{\infty} F_{\lambda} \, d\lambda} \right) + c, 
    \end{aligned}
\end{equation}
with a symbol $c$ to imply the zero-point constant of any value (arbitrary) for the {\it BC} scale, where $\mathcal{F}_{\rm \lambda}$ is monochromatic emergent flux at the wavelength $\lambda$ (ergs cm$^{-2}$ s$^{-1}$) on the surface of the star, $S_{\rm \lambda}$ is the normalized photovisual sensitivity function as given by \citet{Matthews1963}, $f_{\rm \lambda}$ is the monochromatic stellar flux at the Earth (corrected for atmospheric and interstellar extinction) in a relative scale, $F_{\rm \lambda}$ is the absolute monochromatic stellar flux at the Earth (if the star were moved to 10 pc away from the Earth). The terms, which their logarithm was taken, are not different from the luminous efficiency $P$ defined by \citet{Eddington1926}, where $P=\int P(\lambda)I^{'}(\lambda, T_{\rm e})d\lambda/\int I^{'}(\lambda, T_{\rm e})d\lambda$, as in Equation~(\ref{eq:02}). Only the symbols are different here, $P(\lambda)$ is replaced by $S_{\rm \lambda}$ and integrals were taken over black-body fluxes instead of monochromatic intensities. As long as the total radiation output of the star ($L$) is equal to $L=4\pi R^2 \cdot \sigma T^{4}_{\rm e}$ that is, if a star radiates like a black body, all of the four ratios \citep[three above, and the one by][]{Eddington1926} would be equal under the condition that the sensitivity function $S_{\rm \lambda}$ is similar to $P(\lambda)$ because the ratio of the integrated luminous intensity to the total (integrated) intensity would be equal to the ratio of the passing flux to the total (integrated) flux. Therefore, the Equation~(\ref{eq:01}) given by \citet{Eddington1926} defining luminous efficiency is somewhat equivalent to the equations above defining {\it BC} of a star if the constant $c$ were not included in the definition of {\it BC} according to \citet{Heintze1973}. That is, if there were a filter with a transparency profile, which could be expressed by frequencies or wavelengths, similar to the sensitivity profile of a human eye, then the {\it BC} of the star (if $c=0$) would be the logarithm of the number representing integrated luminous efficiency of the human eye multiplied by 2.5. 

Therefore, considering the following comment of \citet{Eddington1926} about Table~\ref{tab:tab1} and the $P$ values,

\begin{pquote}The maximum of $P$ is found to occur at about $T_{\rm eff}=6500^{\rm o}$ so that stars of types F to G  have the greatest luminous efficiency. Presumably, that is because our visual sense has been developed with special reference to sunlight. It is convenient to take the maximum as standard and to define the scale of bolometric magnitude to agree with visual magnitude at this effective temperature. At any other, the temperature $P$ will be smaller, and the star will be brighter bolometrically than visually.
\end{pquote}

One can say that the first and the second paradigms associated with the definition $BC=M_{\rm bol}-M_{\rm V} = m_{\rm bol}-m_{\rm v}$ from \citet{Kuiper1938a} are actually originated from the verse ``At any other temperature $P$ will be smaller and the star will be brighter bolometrically than visually'' contained above comment of Eddington. This verse of Eddington, which was changed later as ``the bolometric magnitude of a star ought to be brighter than its $V$ magnitude'', directly implies both $M_{\rm bol} < M_{\rm V}$, $m_{\rm bol}<m_{\rm V}$ and indirectly implies ``bolometric corrections must always be negative'' ($BC<0$ always).   ($BC<0$ always) is possible mathematically only if both $M_{\rm Bol}<M_{\rm V}$ and $m_{\rm Bol}<m_{\rm v}$. 

Although, the first ($BC<0$ always) and the second ($M_{\rm Bol}<M_{\rm V}$ always) paradigms are initiated by \citet{Eddington1926}, they become unavoidable facts of the first normal science period after \citet{Kuiper1938a}, who added the paradigm 3 by subtracting 0.10 mag from each of his pre-computed {\it BC} values in order to avoid {\it BC} numbers with different signs. The tradition continued by \citet{Lohmann1948, McDonald1952, Eggen1956, Popper1959} and \citet{Wildey1963}. The traditions were raised to levels of paradigms by the time of \citet{Johnson1964, Johnson1966}, who published $BC-T_{\rm eff}$ tables including a few positive {\it BC} values implying that A-F stars have $M_{\rm Bol}>M_{\rm V}$. Not only \citet{Johnson1964, Johnson1966}, there were others such as \citet{Code1976, Flower1977, Flower1996, Bessell1998, Sung2013}, recently \citet{Casagrande2018a, Casagrande2018b} and \citet{Eker2020} who retained limited number of positive {\it BC} in their tables. \citet{Cox2000} was an exception with tabulated {\it BC}, where all {\it BC} values are negative.    

\subsection{The Third Paradigm, the Main Source of Inconsistencies}
\citet{Heintze1973} who inserted arbitrary constant $c$ in Equation (10) commented:

\begin{pquote}For the Sun $F_{\rm \lambda}$ is measured over a sufficient large wavelength region and $BC_{\odot}$ is known from equation $BC=M_{\rm bol}-M_{\rm V}$ and the adopted zero point of the bolometric scale. So, $c$ is known.
\end{pquote}

\noindent Verbal explanation of $c$ after Equation (10) and the above comment is contradictory.  If $c$ is a known quantity, it cannot be arbitrary or vice versa. In fact, {\it BC} producers gave a certain value to it depending upon their private choice. Table~\ref{tab:tab1} indicates $T_{\rm eff}=6000$ K, while \citet{Eddington1926} preferred $T_{\rm eff}=6500$ K (see his comment above), whereas \citet{Kuiper1938a} selected $T_{\rm eff}=6000$ K as the stellar temperature associated with the zero-point constant of the {\it BC} scale as the maximum value of the {\it BC} (+0.10 mag) occurred for him at $T_{\rm eff}=6600$ K. That is, the value $c=0.1$ was subtracted systematically from all {\it BC} in order to adjust the maximum value to zero. \citet{Eggen1956} adopted $c = 0.62$. \citet{Popper1959} took $c=0.58$. Some authors used the solar effective temperature, which could be different among the various authors, while some preferred temperatures of A – F stars while some took $M_{\rm bol,\odot}$  as the zero point, according to Equation~(\ref{eq:08}) above, which may again change according to various authors. \citet{Code1976} declared, ``The choice of a value for the bolometric correction of the Sun is arbitrary (number) and is equivalent to specifying the zero point of the bolometric magnitude scale.''           

No one among the {\it BC} producers said that ``any value of c would do the same job''. Everyone had own reason, which could be different among the different users. \citet{Heintze1973} introduced $c$ just as a constant but not as an arbitrary constant. Otherwise, from minus to plus infinity, any value of $c$ would have been equivalent. However, all the sampled values of $c$ given above (0.1, 0.62, and 0.58) are positive. Thus, they had to be subtracted systematically from {\it BC} to adjust its maximum to be zero.

Nevertheles, considering the definition, $BC=M_{\rm bol}-M_{\rm V} = m_{\rm bol}-V$, of Kuiper, it is not possible for the constant $c$ in Equation~(\ref{eq:10}) to be negative. This is because it has been discussed above that the terms, which their logarithm was taken, are not different from the luminous efficiency $P$ defined by \citet{Eddington1926}. That is the logarithmic terms in Equation~(\ref{eq:10}) are negative or zero as indicated by the numbers in Table~\ref{tab:tab1}, where, at maximum, {\it BC} is expected to be zero. If this is the case, one cannot obtain zero for a {\it BC} by adding two negative numbers. On the other hand, if $L$ and $L_{\rm V}$ are the total (at all wavelengths) and the visual (only the visible part) luminosities of a star represented by its $M_{\rm bol}$ and $M_{\rm V}$, a positive {\it BC} is not allowed by 
\begin{equation}
\label{eq:11}
    BC=M_{\rm bol}-M_{\rm V} = 2.5\log \frac{L_{\rm V}}{L},
\end{equation}
because a positive {\it BC} implies $L_{\rm V}>L$, which is impossible as if a part of a unit were bigger than the unit itself. If both positive and negative numbers are not allowed for $c$ in Equation~(\ref{eq:10}), there is only one number left without any sign; it is zero.  Consequently, it can be concluded that the definition, $BC=M_{\rm bol}-M_{\rm V}=m_{\rm bol}-V$, of Kuiper, and Equation ~(\ref{eq:10}), if there is zero-point constant for {\it BC} values, it must be zero. 

Neither the definition, $BC=M_{\rm bol}-M_{\rm V}=m_{\rm bol}-V$, of \citet{Kuiper1938a}, nor the concept of luminous efficiency ($P$) of \citet{Eddington1926} required a constant. Then, why was \citet{Heintze1973} obliged himself to insert it into his equation? This must be because it was a common practice in these years that the {\it BC} producers had to declare the difference of their {\it BC} values from the previously published ones. Thus, the constant $c$ was there to indicate this difference for readers who compare the different sets of {\it BC} values by different authors.   

Nevertheless, {\it BC} of a star were considered another observable colour like $b-v$, which was the only colour before \citet{Johnson1953} who later introduced the {\it UBV} photometry. This attitude, that is, considering {\it BC} of a star as one of its colours, continued throughout the century; see Table~\ref{tab:tab5}, where {\it BC} in the column before the last were displayed as one of the ten intrinsic colours for calibrating or predicting effective temperatures and/or spectral types of stars. Also covering these tables, Astronomers’ handbook, ``Allen’s Astrophysical Quantities'' \citep{Cox2000} confirms this attitude also.

A consistent hint why should be a constant in Equation~(\ref{eq:10}), came from \citet{Pecker1973}, who declared:   
\begin{equation}
\label{eq:12}
M_{\rm bol}=-2.5\log L + C,
\end{equation}
which is the basic rule of the magnitude system revealing how $L$ of a star is related to its absolute bolometric magnitude $M_{\rm bol}$. It is clear that if the constant $C$ is equal to $+2.5\log L$, then $M_{\rm bol}=0.0$. Therefore, the constant $C$ in Equation~(\ref{eq:12}) is called the zero-point constant. $m=-2.5\log \left(f/f_{\rm 0}\right)$ is given in Allen’s Astrophysical Quantities \citep{Cox2000} when describing apparent magnitudes of photometric systems, where $f$ is the measured flux (corrected for atmospheric effects), and $f_{\rm 0}$ is the corresponding flux for a star of zero apparent magnitude ($m=0$). It is clear that if $f=f_{\rm 0}$, then $m=0$. Since $f_{\rm 0}$ is a constant, the formula $m=-2.5\log \left(f/f_{\rm 0}\right)$ could be written in the same format as:
\begin{equation}
\label{eq:13}
m_{\rm bol}=-2.5\log f + c,
\end{equation}
where the constant $c$ is $2.5\log f_{\rm 0}$. Because an absolute magnitude is directly related to the $L$ of a star, but the apparent magnitude is related to the flux coming to the eye of the observer from the same star, the zero-point constants associated with $L$ and $f$ are not the same. The difference is indicated by capitalization: $C$ for the absolute, $c$ for the apparent.
  
To understand what happens, let us assume that there is no atmosphere. Thus, the total flux ($f$) at all wavelengths will come to the eye of the observer and to the detector of the telescope with a blue-sensitive photographic plate from the star. Eye and the detector will sense visual and relatively bluer photons, Equation~(\ref{eq:13}) will transform to $V=-2.5\log f_{\rm V}+c_{\rm V}$ and  $B=-2.5\log f_{\rm B}+c_{\rm B}$ for the eye and the detector respectfully. Using relative photometry and one reference star with a known or assigned brightness such as $V_{0}$ or $B_{\rm 0}$, $B$ and $V$ magnitudes of all stars in the sky could be determined without knowing individual flux values ($f_{\rm V}$ and $f_{\rm B}$) and the zero-point constants ($c_{\rm V}$ and $c_{\rm B}$) this is because when determining apparent magnitude difference of two stars, the zero-point constants cancel nicely at the first step as 
\begin{equation}\label{eq:14}
    \begin{aligned}
        V(2) - V(1) &= -2.5 \log \left( \frac{f_V (2)}{f_V (1)} \right), \\
        \text{while} \quad B(2) - B(1) &= -2.5 \log \left( \frac{f_B (2)}{f_B (1)} \right),
    \end{aligned}
\end{equation}
if comparisons are made staying in the same optical band. This is just because the magnitude difference of the two stars is equal to the ratio of the received fluxes. If all stars are compared for the second step, one only needs to know the flux received from the reference star and its magnitude to determine the relative brightness (magnitudes) of all stars. Changing the wavelength sensitivity of the detector and/or changing the filter of the telescope, different sorts of magnitude scales with their special zero-point constants could be determined. Each one of the various magnitude scales has the same magnitude units but different zero-points, like temperature scales Celsius and Kelvin. Colours were invented as different kinds of magnitude scales with their own zero-point constants to indicate effective temperatures of stars, not for one to compare magnitudes of stars at two different parts of the stellar spectrum. For example, the colours $U-B$, $B-V$, $V-R$, ..., displayed in Table~\ref{tab:tab5}, were set to be zero for a hypothetical star of A0V, in practice, Vega. This, too, is also allowed in the relative photometry that one does not need to know the zero-point constants of the magnitude scales involved (for example, zero points of $V$ and $B$ for the colour $B-V$) to determine the intrinsic colours of nearby stars with no interstellar extinction in a similar manner by determining magnitude difference of two stars.

Vega appears to be the reference star also for the intrinsic colours. That is, all the colours in Table~\ref{tab:tab5} are zero for Vega. Note that the colours of the reference star were assumed to be zero but not its apparent magnitude. $V=0.03$ mag has been measured by \citet{Johnson1966} and \citet{Bessell1998}. The standard $V=0.03$ mag is cited by \citet{Bohlin2014, Cox2000, Girardi2002, Bessell2012, Casagrande2014}. For the other bands, Vega is found to be just slightly different ($\sim 0.02$ mag) at most bands \citep{Rieke2008}.

Relative photometry not only permits one to determine apparent magnitudes but also absolute magnitudes of all stars in the sky if their parallaxes were measured and if their interstellar extinctions were also known without knowing the zero-point constants of the absolute magnitudes, that is, without knowing the capitalised $C$ associated with individual photometric bands.  

All of the advantages of relative photometry vanish if one tries to determine the bolometric magnitudes of the stars and the bolometric corrections appearing like the other colours in Table~\ref{tab:tab5}. This is because there is no telescope nor a detector to observe a star at all wavelengths. Moreover, unlike the other colours, which are special magnitude scales indicating effective temperatures, not for showing the difference of the two magnitudes fixed to be zero artificially using Vega, the {\it BC} values in Table~\ref{tab:tab5} are the numbers indicating the differences between the total luminosity (or total flux) and the partial luminosity (or received flux) of a star constrained within a wavelength range proper to an eye or a filter. Therefore, it is not possible to bypass the zero-point constants of the bolometric magnitudes specified by Equations~(\ref{eq:12}) and ~(\ref{eq:13}) nor the ones involving with the visual band, which could be symbolised $C_{\rm V}$ for the absolute, $c_{\rm V}$ for the apparent visual magnitudes. Having realised these facts, \citet{Code1976} suggested two new formulations 
\begin{align}
    B.C. &= -2.5 \log \left( \int_{0}^{\infty} f_{\lambda} \, d\lambda \right) - V + C_1, \label{eq:15}
    \\[10pt]
    B.C. &= 2.5 \log \left( \frac{\int_{0}^{\infty} S_{\lambda} f_{\lambda} \, d\lambda}{\int_{0}^{\infty} f_{\lambda} \, d\lambda} \right) + C_2, \label{eq:16}
\end{align}
for calculating {\it BC} of a star according to the original definition, $B.C.=m_{\rm bol}-V$, of \citet{Kuiper1938a}, where $m_{\rm bol}$ and $V$ are the apparent bolometric and visual magnitudes, $f_{\rm \lambda}$ is the flux per unit wavelength interval received outside the atmosphere from a star at wavelength $\lambda$, $S_{\rm \lambda}$ is the sensitivity function of the $V$ magnitude system and $C_1$ and $C_2$ are just constants. \citet{Code1976} commented ``Equations~(\ref{eq:15}) and ~(\ref{eq:16}) are both completely equivalent to Equation~(\ref{eq:07}) and while Equation~(\ref{eq:16}) may be used to determine bolometric corrections for any SED irrespective of the units in which $f_{\rm \lambda}$ is expressed, Equation~(\ref{eq:15}) can be applied only when $f_{\rm \lambda}$ is in the units for which the constant $C_1$ has been determined''. Readers must pay attention here: neither of the two constants ($C_1$ and $C_2$) were claimed to be arbitrary. 

Using the sensitivity function $S_{\rm V}$, outside the atmosphere, from  \citet{Matthews1963} and the Solar SED of \citet{Arvesen1969}, \citet{Code1976} found $C_2=0.958$. while $C_2=0.946$ was calculated by them from the SED of the Sun by \citet{Labs1968}. Consequently, the uncertainty of $C_2$ was estimated at 0.01 mag. \citet{Code1976} also estimated the zero-point constant of the visual apparent magnitudes $c_{\rm V}=-12.47$ from the mean of 21 stars with observationally estimated SEDs using the formula:
\begin{equation}\label{eq:17}
V = -2.5 \log \left( \int_{0}^{\infty} S_{\lambda} f_{\lambda} \, d\lambda \right) + c_V.
\end{equation}

Despite $c_{\rm V} = -11.48\pm0.03$ mag was also found by \citet{Code1976} from the apparent magnitude $V=-26.74\pm 0.03$ mag \citep{Johnson1965, Code1973} and the total flux of the Sun, $f=1.360\pm0.014\times 10^6$ ergs cm$^{-2}$ s$^{-1}$ \citep{Duncan1969, Allen1973}, and $c_{\rm V}=-12.47$ mag were preferred to calculate the constant in Equation~(\ref{eq:15}) from $C_1=-12.47+0.959=-11.51$ mag. It is important to note that a positive value assigned to $C_2$ invalidates the previous arguments on Equation~(\ref{eq:11}), from which the first and the second paradigms originated. For advocating the third paradigm, ``the zero point of bolometric corrections are arbitrary'', \citet{Hayes1978} expressed $m_{\rm bol}$ with apparent magnitudes from the definition of Kuiper as
\begin{equation}
\label{eq:18}
    m_{\rm bol} = -2.5\log_{10} f + C = V+B.C., 
\end{equation}
and
\begin{equation}
\label{eq:19}
    m_{\rm bol,*} -m_{\rm Bol,\odot}= -2.5\log_{10} (f_*/f_{\odot}). 
\end{equation}
argued ``We see from Equations~(\ref{eq:18}) and ~(\ref{eq:19}) that the zero-point of {\it BC} scale is arbitrary and that a {\it BC} scale may be measured by measuring $f$ and $V$ for the suitable number of stars''. It must have been because of this comment of Hayes that many astronomers and astrophysicists, even today, may think the zero-point of the {\it BC} scale is arbitrary despite Hayes had already corrected himself by removing the statement ``the zero point of {\it BC} scale is arbitrary'' from the manuscript of the next paper \citep{Hayes1985}. Is not it paradoxical to say, ``the zero-point of {\it BC} scale is arbitrary” and then to say, “{\it BC} scale may be measured by measuring $f$ and $V$ for the suitable number of stars''? If something is a measured quantity, it is not arbitrary.

At last, \citet{Torres2010} argued, ``The bolometric correction is usually defined as the quantity to be added to the apparent magnitude in a specific passband (in the absence of interstellar extinction) to account for the flux outside the band''. Referring to $BC_V=m_{\rm bol}-V=M_{\rm bol}-M_{\rm V}$, which is the definition of \citet{Kuiper1938a, Torres2010} continued ``Note that this definition is usually interpreted to imply that the bolometric corrections must always be negative...'', (paradigm one), ``... although many of the currently used tables of empirical $BC_{\rm V}$ values violates this condition'' while he addressed the second and the third paradigms as:

\begin{pquote}
The scale adopted by \citet{Flower1996} is such that $BC_{\rm V,\odot}=-0.080$. As a result, his bolometric corrections for stars between $T_{\rm eff}\approx 6400$ K and $T_{\rm eff}\approx 8500$ K (spectral types approximately F5 to A5) are positive, seemingly conflicting the idea that the bolometric magnitudes ought to be brighter than $V$ magnitudes, according to $BC_V=m_{\rm bol}-V=M_{\rm bol}-M_{\rm V}$. Other tables with a similar zero point as Flower’s share the same problem. In reality, however, the contradiction is of no consequence because of the arbitrary nature of the zero point. Luminosities inferred for stars are never affected if a consistent value of $M_{\rm bol,\odot}$ is used.
\end{pquote}

After expressing the definition, $BC_V=m_{\rm bol}-V$, in its integral form as 
\begin{equation}
\label{eq:20}
BC_{\rm V} = 2.5 \log \left( \frac{\int_{0}^{\infty} S_{\lambda} (V) f_{\lambda} \, d\lambda}{\int_{0}^{\infty} f_{\lambda} \, d\lambda} \right) + C_2.
\end{equation}
\citet{Torres2010} asserted ``the constant $C_2$ contains an arbitrary zero point that has been a common source of confusion. This zero point has traditionally been set using the Sun as the reference'' as if between the lines he reminded \citet{Hayes1978}, but not \citet{Hayes1985}, who corrected himself regarding the zero-point constant of the {\it BC} scale. After he complained by saying ``While the zero point of $BC_{\rm V}$ is completely arbitrary, and no particular scale has been officially endorsed by the International Astronomical Union (IAU), it is common for some authors of these tabulations to define the scale by adopting a certain value for  $BC_{\rm V,\odot}$, sometimes for historical reasons'', Torres declared:

\begin{pquote}
Somewhat surprisingly, the IAU has not issued a formal resolution on the matter of $BC_{\rm V}$ zero points, although two of its Commissions did agree at the Kyoto meeting of 1997 \citep[][pp. 141 and 181]{Andersen1999} on a preferred scale that is equivalent to adopting a value for $M_{\rm Bol,\odot}$\footnotemark. The scale was set by defining a star with $M_{\rm bol}=0.00$ to have an absolute radiative luminosity of $L=3.055\times 10^{28}$ W \citep[see also,][]{Cayrel2002}. The rationale was that this value together with the nominal bolometric luminosity of the Sun adopted by the international Global Oscillation Network Group project ($L_{\odot}=3.846\times 10^{26}$ W, according to the IAU Commission reports cited above) leads exactly to $M_{\rm bol,\odot}=4.75$, which is the bolometric magnitude for the Sun listed in the 1973 edition of Astrophysical Quantities \citep{Allen1973}. This was a widely used source at the time (and still is, by some), so it was thought to be a logical choice. Effectively, therefore, the scale is set by this value of $M_{\rm bol,\odot}$. Combined with our adopted solar brightness of $V_{\odot}=-26.76$, it implies $BC_{\rm V,\odot}=-0.06$. As it turns out, however, the most recent edition of Astrophysical Quantities \citep{Cox2000} did not follow that recommendation and adopted a slightly different zero point. ... the adoption of a value of $BC_{\rm V,\odot}$ or a value of $M_{\rm bol,\odot}$ may not be the most convenient way to solve the immediate problem faced by users
\end{pquote}
\footnotetext{This scale was also adopted by IAU Division IV at the 2003 meeting of the IAU in Sydney \textcolor{blue}{(Engvold 2007)}.}
The problems with the definition of {\it BC} \citep{Kuiper1938a} were summarized by \citet{Torres2010} without mentioning the word ``paradigm'' together with his opinions on the possible solutions, which could be initiated by IAU. 

\section{IAU 2015 General Assemly Resolution B2, New Era of B.C.}

There were two basic motivations for IAU to resolve the problems associated with bolometric magnitudes and bolometric corrections. 1) A strong need for standardized absolute and apparent bolometric magnitude scales, and 2) to stop problems originating from multiple zero points of bolometric corrections. Thus, the following comments were issued before the final statements of the resolution.

\begin{pquote}1. The need for a standardized absolute and apparent bolometric magnitude scale for accurately and repeatably transforming photometric measurements into radiative luminosities and irradiances, independently of the variable Sun,\\
2. that multiple zero points for bolometric corrections pervade the literature due to the lack of a commonly adopted standard zero point for the bolometric magnitude scale
\end{pquote}

The most logical and efficient way was chosen in contrast to \citet{Torres2010}, who expected a formal resolution on the matter of $BC_{\rm V}$ zero points. On the contrary, IAU 2015 General Assembly Resolution B2 (after IAU 2015 GAR B2) was issued to fix the zero point of the absolute bolometric magnitude scale according to the direct relation between $M_{\rm bol}$ and $L$ of the stars as \citet{Pecker1973} recommended (see Equation~(\ref{eq:12}) above). The resolution is 

\begin{pquote}A radiation source with absolute bolometric magnitude $M_{\rm Bol}=0$ has a radiative luminosity of exactly $L_{0}=3.0128\times 10^{28}$ W and the absolute bolometric magnitude $M_{\rm Bol}$ for a source of luminosity $L$ (in W) is
\begin{equation}
    \label{eq:21}
    M_{\rm Bol} = -2.5 \log \left( \frac{L}{L_{0}} \right) = -2.5 \log L + 71.197425 ...
\end{equation}
the zero point was selected so that the nominal solar luminosity corresponds closely to absolute bolometric magnitude $M_{\rm Bol, \odot} = 4.74$ mag, the value most commonly adopted in the recent literature \citep[e.g.][]{Bessell1998, Cox2000, Torres2010}\footnotemark.
\end{pquote}
\footnotetext{\url{https://www.iau.org/static/resolutions/IAU2015_English.pdf}} \noindent declared together with a reminder: the notations of $M_{\rm Bol}$ and $m_{\rm Bol}$ were adopted by Commission 3 (Notations) at the 6th IAU General Assembly in Stockholm in 1938. It is declared that $M_{\rm Bol}$ and $m_{\rm Bol}$ refer specifically to absolute and apparent bolometric magnitudes, respectively. Thus, the zero-point constant of the bolometric magnitude scale is $C_{\rm Bol} = 71.197 425 ...$ if $L$ is in SI units, and $C_{\rm Bol} = 88.697 425 ...$ if $L$ is in cgs units. The zero point of apparent bolometric magnitudes was also fixed similarly as 

\begin{pquote}Apparent bolometric magnitude $m_{\rm Bol}=0$ mag corresponds to an irradiance or heat flux density of $f_{0}=2.518 021 002 ...10^{-8}$ W m$^{-2}$ and hence the apparent bolometric magnitude $m_{\rm Bol}$ for an irradiance $f$ (in W m$^{-2}$) is
    \begin{equation}
    \label{eq:22}
m_{\rm Bol}=-2.5\log(f⁄f_{0})=-2.5\log f-18.997 351 ...
    \end{equation}
\end{pquote}

At first glance, Equations~(\ref{eq:21}) and~(\ref{eq:22}) may appear independent, but they are not because the irradiance $f_0$ corresponds to a measurement from an isotopically emitting radiation source with absolute bolometric magnitude $M_{\rm Bol}= 0.0$ mag (luminosity $L_{0}$) at the standard distance of 10 pc (based on the IAU 2012 definition of the astronomical unit). 
\vspace*{-10pt}

\subsection{Vega System of Magnitudes and Multi-band {\it BC}}
One possible counterargument advocating the arbitrariness of the {\it BC} scale and, consequently, the other two paradigms is that IAU 2015 GAR B2 did not set the bolometric correction scale. It defined only the bolometric magnitude scale set to SI irradiance ($m_{\rm Bol}$, $f_{\rm Bol}$) and luminosity ($M_{\rm Bol}$, $L_{\rm Bol}$) values, whereas the photometric magnitudes are independently defined by standard stars (for Vega system magnitudes, or fluxes for the ST or AB scale). The IAU could easily set the former, but not the latter, such as the Johnson $V$ system. Because the IAU did not set the {\it BC} scale, nothing new could be claimed. A firm knowledge such as the arbitrariness of the {\it BC} scale, which has been actively used at least for about eight decades, cannot be changed. 

First, it is not true to think that the arbitrariness of the {\it BC} scale has been a firm knowledge for about eight decades. As was explained above, there was only one author \citep{Hayes1978} who claimed the zero point of {\it BC} scale is arbitrary before \citet{Torres2010}. This author, \citet{Hayes1978}, too, corrected himself by removing the statement ``the zero-point of {\it BC} scale is arbitrary'' from the discussions of the same Equations~(\ref{eq:18}),~(\ref{eq:19}) in the new paper \citep{Hayes1985}. The unknown constants in Equations such ~(\ref{eq:10}), ~(\ref{eq:12}), ~(\ref{eq:13}), ~(\ref{eq:15})-(\ref{eq:18}), ~(\ref{eq:20}) were interpreted as arbitrary by \citet{Torres2010} and the others who are heavily influenced by the three paradigms related to {\it BC}. 

Second, from the original definition of \citet{Kuiper1938a}, Equation~(\ref{eq:18}) above, from \citet{Hayes1978, Hayes1985} could be adapted to any photometric system: e.g., for the Johnson system of apparent magnitudes ($U, B, V, R, I$) as
\begin{equation}\label{eq:23}
    \begin{aligned}
           m_{\rm Bol} &= \xi+BC_{\xi}=U+BC_{\rm U}=B+BC_{\rm B}=V+BC_{\rm V} = R+BC_{\rm R}=I+BC_{\rm I},  
    \end{aligned}
\end{equation}
and for the absolute magnitudes as
\begin{equation}\label{eq:24}
    \begin{aligned}
        M_{\rm Bol} &= M_{\xi}+BC_{\xi}=M_{\rm U}+BC_{\rm U}=M_{\rm B}+BC_{\rm B} = M_{\rm V}+BC_{\rm V}=M_{\rm R}+BC_{\rm R}=M_{\rm I}+BC_{\rm I},
    \end{aligned}
\end{equation}
where the subscripts indicate a filter in a photometric system. It would have been illogical to set innumerable zero-points for the {\it BC} of each band ($BC_{\xi}$) in various photometric systems while an easy way exists to resolve the problem in a single step, as IAU 2015 GAR B2 did. It is very obvious that fixing the zero-point constant of absolute bolometric magnitudes as in Equation~(\ref{eq:21}) automatically fixes all of the zero-points of the {\it BC} scales of various bands of various photometric systems including the zero-points of apparent bolometric and the corresponding apparent magnitudes.

IAU 2015 GAR B2 is a revolutionary document not only because it initiated the steps of the three paradigms, ``bolometric corrections must always be negative'', ``the bolometric magnitude of a star ought to be brighter than its $V$ magnitude'' and ``the zero point of bolometric corrections are arbitrary'' are wrong and inconsistent \citep{Eker2022}, but also because it removes problematic classical equation,
\begin{equation}
   \label{eq:25}
M_{\rm Bol}=M_{\rm Bol,\odot}-2.5\log(L⁄L_{\odot}).
\end{equation}
from the classical steps for calculating the absolute magnitude of a star from its observables: $R$ and $T_{\rm eff}$. 

Before, and even after, IAU 2015 GAR B2, Equation~(\ref{eq:25}) was claimed to be the main source of the arbitrariness attributed to the zero points of bolometric magnitudes and corrections \citep{Hayes1978, Torres2010, Eker2021a, Eker2021b, Casagrande2018a, Casagrande2018b, Andrae2018}. The Solar luminosity varies according to \citet{Frohlich1998}, with an amplitude about $1.5\times 10^{-3}$ ($\approx 0.004$ mag) because of magnetic activity. Therefore, Equation~(\ref{eq:25}) cannot be the fundamental relation to calculate absolute bolometric magnitudes of stars. See the first motivation of IAU for the resolution, IAU 2015 GAR B, where avoiding the variable the Sun stated clearly. In another saying, it was not the choice of $M_{\rm Bol,\odot}$ and the luminosity of the Sun ($L_{\odot}$) to determine the zero-point constant of absolute bolometric magnitude scale which is $C_{\rm Bol} = 71.197 425 ...$ if $L$ is in SI units, and $C_{\rm Bol} = 88.697 425 ...$ if $L$ is in cgs units. On the contrary, it was the choice of $C_{\rm Bol} = 71.197 425 ...$ and the best estimate of solar luminosity $L_{\odot}= 4\pi(1 \text{ au})^{2} S_{\odot} = 3.8275\pm0.0014\times 10^{26}$ W, which is a value computed from the recently estimated best value of the solar constant $S_{\odot} = 1361\pm 1$ W m$^{-2}$, obtained from the total solar irradiance (TSI) data of 35 years of space-born observations during the last three solar cycles \citep{Kopp2014}, which determined the absolute bolometric magnitude of the Sun as $M_{\rm Bol,\odot}=4.739 996 ...$ mag. Therefore, $M_{\rm Bol,\odot}=4.74$ mag and $L_{\odot}=3.828\times 10^{26}$ W were suggested as the nominal solar absolute bolometric brightness and the nominal solar luminosity for the ones who still want to use Equation~(\ref{eq:25}).

\section{Discussions}
 \subsection{New Definitions and Concepts}
There are two different definitions of bolometric corrections. Both are still functioning. The one which was suggested by \citet{Kuiper1938a} as 
$B.C.=M_{\rm bol}-M_{\rm pv}=m_{\rm bol}-m_{\rm pv}$, listed as ``$BC=m_{\rm bol}-V$ (always negative)'' in the handbook of astronomers in page 381, Allen’s Astrophysical Quantities \citep{Cox2000}, but could be adapted for any passband ($\xi$) in any photometric system as  
$BC_{\xi}=M_{\rm bol}-M_{\xi}=m_{\rm bol}-\xi$, therefore, the very first definition could be expressed verbally as ``the difference between the bolometric and apparent magnitudes of a star''. The second and relatively more modern definition is ``the quantity to be added to the apparent magnitude in a specific passband (in the absence of interstellar extinction) in order to account for the flux outside the band'' as given by \citet{Torres2010}.

Both of the definitions could be considered correct but incomplete. {\it BC} of a star is better to be defined as ``{\it BC} of a star is the difference between its bolometric and one of its apparent magnitudes in their absolute scales, that is, without ignoring both of the zero-point constants associated with bolometric and filtered quantities, unlike the other colours operating in relative scale thus the difference between the two apparent (or absolute) magnitudes were conventionally assumed to be zero for a hypothetical star of the spectral type A0V''.

IAU 2015 GAR B2 declared Equation~(\ref{eq:21}) as the direct relation between the absolute bolometric magnitude of a star ($M_{\rm Bol}$) and its luminosity ($L$). $M_{\rm Bol}=M_{\rm Bol,\odot}-2.5\log (L/L_{\odot})$, the classical formula is cancelled and replaced by Equation~(\ref{eq:21}). Knowing $L$ of a star, its $M_{\rm Bol}$, or knowing $M_{\rm Bol}$, its $L$ can be calculated directly. Equation~(\ref{eq:21}) could be adopted for any passband of any photometric system, such as:
\begin{equation}
   \label{eq:26}
M_{\xi}=-2.5\log L_{\xi}+C_{\xi},
\end{equation}
where $L_{\xi}$ is the partial luminosity, and $C_{\xi}$ is the zero-point constant of the magnitude scale for the photometric band $\xi$. Unlike, $C_{\rm Bol}=71.197 425 ...$ if $L$ is in SI units for the bolometric magnitudes, $C_{\xi}$ may not be found in literature because astronomers commonly prefer relative photometry, and one can do relative photometry without knowing the zero-point constant. Therefore, $C_{\xi}$ is left as an unknown in this study, but it is a definite value, as accurate as the photometric system practised.

IAU 2015 GAR B2 declared Equation~(\ref{eq:22}) as the direct relation between the apparent bolometric magnitude of a star and its irradiance or the heat flux density, assuming no extinctions. We can adopt a similar relation between the apparent magnitude ($\xi$) of a star at any passband in a photometric system and heat flux density (irradiance) received from the star limited by the wavelength range of the passband as 
\begin{equation}
   \label{eq:27}
\xi=-2.5\log f_{\xi}+c_{\xi},
\end{equation}
where $f_{\xi}$ is the heat flux density (irradiance) within the wavelength range of the pass band if there is no extinction and $c_{\xi}$ is the zero-point constant of apparent magnitudes for the passband chosen. The relation between $L_{\xi}$ and $f_{\xi}$ and the relation between $C_{\xi}$ and $c_{\xi}$ are not independent. A similar relationship between $L$ and $f$ must also exist between $L_{\xi}$ and $f_{\xi}$. A similar relationship between $C_{\rm Bol}$ and $c_{\rm Bol}$ must also exist between $C_{\xi}$ and $c_{\xi}$. Unlike Equation~(\ref{eq:22}) where $c_{\rm Bol}=-18.997 351 ...$ corresponding to the heat flux density $f_{0}=2.518 021 002 ... 10^{-8}$ W m$^{-2}$, $c_{\xi}$ may not be found in the literature because astronomers commonly use relative photometry. Therefore, $c_{\xi}$ is left as an unknown in this study, but it must be a definite value, as accurate as the photometric system practised. Note that Equation~(\ref{eq:27}) is the generalized form of Equation~(\ref{eq:22}), which is valid for apparent visual magnitudes. 

One can obtain BC of a star for the passband $\xi$ according to the generalized form with absolute magnitudes ($BC_{\xi}=M_{\rm Bol}-M_{\xi}$) by subtracting Equation~(\ref{eq:26}) from Equation~(\ref{eq:21}) as:
\begin{equation}
 \label{eq:28}
BC_{\xi} = M_{\rm Bol} - M_{\xi} = 2.5 \log \left( \frac{L_{\xi}}{L} \right) + (C_{\rm Bol} - C_{\xi}).
\end{equation}
while the generalized form with apparent magnitudes ($BC_{\xi}=m_{\rm Bol}-\xi$) by subtracting Equation~(\ref{eq:27}) from Equation~(\ref{eq:22}) as:
\begin{equation}
 \label{eq:29}
BC_{\xi} = m_{\rm Bol} - \xi = 2.5 \log \left( \frac{f_{\xi}}{f} \right) + (c_{\rm Bol} - c_{\xi}).
\end{equation}

Combining Equation~\ref{eq:28} and Equation~\ref{eq:29}, 
\begin{equation}\label{eq:30}
    \begin{aligned}
        BC_{\xi} &= 2.5 \log \left( \frac{L_{\xi}}{L} \right) + (C_{\rm Bol} - C_{\xi}) = 2.5 \log \left( \frac{f_{\xi}}{f} \right) + (c_{\rm Bol} - c_{\xi}) = 2.5 \log \left( \frac{\int_{0}^{\infty} S_{\lambda} (\xi) f_{\lambda} \, d\lambda}{\int_{0}^{\infty} f_{\lambda} \, d\lambda} \right) + C_2 (\xi),
    \end{aligned}
\end{equation}
is the formula to calculate {\it BC} of the star according to the new definition of {\it BC} for the passband $\xi$, where $C_2(\xi)=C_{\rm Bol}-C_{\xi}=c_{\rm Bol}-c_{\xi}$ is the zero point constant of the {\it BC} scale for passband $\xi$. The part after the first equal sign is the {\it BC} from the absolute bolometric magnitudes, which indicates the $BC_{\xi}$ is related to the fractional luminosity ($L_{\xi}⁄L$) within the passband $\xi$. The part after the second equal sign is the {\it BC} from the apparent magnitudes, which shows $BC_{\xi}$ is also related to the fraction of the heat flux ($f_{\xi}⁄f$) passing through the filter $\xi$ above the atmosphere if there is no extinction. The last part of the equation is an integral form of the fractional heat flux, which allows one to calculate $BC_{\xi}$ numerically if the SED of the star and transition profile ($S_{\lambda}(\xi)$) of the filter is known. Equality of ~(\ref{eq:28}) and~(\ref{eq:29}) depends on the equality of $C_{\rm Bol}-C_{\xi}$ to $c_{\rm Bol}-c_{\xi}$, which was already confirmed \citep{Eker2022}, and on the assumption that radiation is isotropic, thus, $L=4\pi d^{2} \cdot f$ and $L_{\xi}=4\pi d^{2} \cdot f_{\xi}$, where $d$ is the distance of the star.

\subsection{Replacements}
According to \citet{Kuhn1970}, the next normal science period does not start until the misleading paradigms of the previous period are replaced by the new ones. We are hoping that the replacements offered in this study are sufficiently convincing that the paradigms of the previous period will be totally discarded.

The third paradigm, ``the zero point of bolometric corrections are arbitrary, '' is clearly stated and advocated by \citet{Torres2010}. However, according to the new definition and the new formula for $BC_{\xi}$, the zero-point constant $C_2(\xi)$ is a firm constant with accuracy, which could be determined by observations. Subscript 2 indicates that it is set by the difference between the two zero-point constants as indicated by Equations~(\ref{eq:28}) and~(\ref{eq:29}). Furthermore, it must have a positive value because: Let us assume a star has $M_{\rm Bol}= 0.0$, then according to Equation~(\ref{eq:21}), $C_{\rm Bol} = 71.197 425 ...$ if $L$ is in SI units, and $C_{\rm Bol} = 88.697 425 ...$ if $L$ is in cgs units. On the other hand, Equation~(\ref{eq:27}) implies $M_{\xi}=0.0$ for the same star if $BC_{\xi}$ of this star is zero. Then, $L_{\xi}<L$, which is true for all stars, implies $C_{\xi}<C$ and both are positive. Lastly, Equations~(\ref{eq:27}) and~(\ref{eq:30}) indicates $C_{2}(\xi)=C_{\rm Bol}-C_{\xi}$ is a positive number. 

\citet{Code1976} determined it for the visual band as $C_2(V)=0.958\pm0.010$ mag using observed SEDs of 21 nearby stars. We are aware of an official two-year project (TÜBİTAK 123C161), which aims to determine the zero-point constant of {\it BC}, primarily, for the Johnson $V$ filter and, if possible, for the $B$ filter (\textcolor{blue}{Yücel, 2025})\footnote{private communications} from the numerous high-resolution spectra of stars distributed all over the H-R diagram is ongoing. Therefore, the third paradigm must be discarded with a firm knowledge that ``the zero point constants of bolometric corrections are not arbitrary. It could be determined observationally as accurate as up to the sixth digits after the decimal, that is, it could be in the limit as accurate as $C_{\rm Bol}$ given in Equation~(\ref{eq:21})''.  

According to Equation~(\ref{eq:30}), it is not that the bolometric correction itself ($BC_{\xi}$), but the bolometric correction minus the zero-point constant [$BC_{\xi}$ - $C_2(\xi)$] is always less than zero. Therefore, this must be the second replacement, we suggest, for discarding the first paradigm, ``bolometric corrections must always be negative''. This is because, Equation~(\ref{eq:30}) could be re-written as
\begin{equation}\label{eq:31}
    \begin{aligned}
        BC_{\xi} - C_2 (\xi) &= 2.5 \log \left( \frac{L_{\xi}}{L} \right) = 2.5 \log \left( \frac{f_{\xi}}{f} \right) = 2.5 \log \left( \frac{\int_{0}^{\infty} S_{\lambda} (\xi) f_{\lambda} \, d\lambda}{\int_{0}^{\infty} f_{\lambda} \, d\lambda} \right),
    \end{aligned}
\end{equation}
where it is clear that the fractional luminosity ($L_{\xi}/L$), fractional flux ($f_{\xi}/f$) and integral form of the fractional flux 
($\int_{0}^{\infty}S_{\lambda}(\xi)f_{\lambda}d\lambda/\int_{0}^{\infty} f_{\lambda} d\lambda$)
are the same number, which is positive and less than one. The Logarithm of a number which is positive and less than one is a negative number. Therefore, $BC_{\xi}-C_2(\xi)<0$ always. 

It can be written from Equation~(\ref{eq:31}) that
\begin{equation}\label{eq:32}
    \frac{L_{\xi}}{L} = 10^{\textstyle \frac{BC_{\xi} - C_2 (\xi)}{2.5}}.
\end{equation}
which indicates the fractional luminosity of a star within the wavelength range covering the filter $\xi$ could be computed if its {\it BC} and corresponding zero-point constant are known. Similarly, it can be written from Equation~(\ref{eq:31}) that 
\begin{equation}\label{eq:33}
    \frac{f_{\xi}}{f} = 10^{\textstyle \frac{BC_{\xi} - C_2 (\xi)}{2.5}}.
\end{equation}
which indicates fractional luminosity ($L_{\xi}/L$) of a star is equal to its fractional heat flux ($f_{\xi}/f$) at above the Earth’s atmosphere if there is no interstellar extinction. 

Because the zero-point constant, $C_2(\xi)$, is a positive number, as discussed, not only $BC_{\xi}<0$ but also $0<BC_{\xi}<C_2(\xi)$ are valid to produce a consistent fractional luminosity according to Equation~(\ref{eq:32}) and a true fractional flux according to Equation~(\ref{eq:33}). Only if $BC_{\xi}\geq C_2(\xi)$, invalid consequences are inevitable. This is because, if $BC_{\xi}\geq C_2({\xi})$, Equations~(\ref{eq:32}) and ~(\ref{eq:33}) imply $L_{\xi}\geq L$ and $f_{\xi}\geq f$, which are unphysical. Therefore, the statement (the second paradigm) ``the bolometric magnitude of a star ought to be brighter than its $V$ magnitude'' cannot be generalized to all stars even if there is only one star with a positive $BC_{\xi}$. We do not have a replacement for the second paradigm, so it must be forgotten. 

\subsection{Standardization and Future Research}

The primary purpose of IAU 2015 GAR B2 was to stop producing inconsistent bolometric corrections and ensure standardization. The standard bolometric corrections were defined verbally by \citet{Eker2021a}, while \citet{Eker2021b} defined the standard stellar luminosities. If a {\it BC} is produced using observed $R$ and $T_{\rm eff}$ of a star, giving its $L$ first, and then by using the equation $M_{\rm Bol} = -2.5 \log L + 71.197425...$ (if $L$ is in SI units) as proposed by IAU 2015 GAR B2, and $ BC_{\xi} = M_{\rm Bol} - M_{\xi}$ to include its absolute $\xi$ magnitude free from extinctions, the produced $BC_{\xi}$ are called standard. If $M_{\rm Bol}$ of a star is obtained directly from a pre-computed standard $BC_{\xi}$ and $M_{\xi}$ first, then its $L$ is computed using $M_{\rm Bol} = -2.5 \log L + 71.197425...$ (if $L$ is in SI units) as proposed by IAU 2015 GAR B2, the obtained $L$ is called standard luminosity.

Accuracy of standard luminosities in the era after {\it Gaia} were also studied \citep{Eker2021b}. It has been found that the accuracy of standard luminosities is comparable to the accuracy of the luminosities from the direct method with the Stefan-Boltzmann law, $L=4\pi R^2 \cdot \sigma T_{\rm eff}^4$. Using multiband bolometric corrections at Johnson {\it B, V}, and {\it Gaia} $G$, $G_{\rm BP}$, $G_{\rm RP}$ filters of 209 double-lined detached eclipsing binaries having main-sequence components and {\it Gaia} EDR3 parallaxes, \citet{Bakis2022} found a new method to improve accuracy of the standard luminosities much better ($\approx$3 times) than the accuracy of the luminosities from the classical method with the Stefan-Boltzmann law. The new method, which uses multiband bolometric corrections, was tested by \citet{Eker2023} by recovering the luminosities and radii of 341 host stars with the most accurate effective temperatures. The error histograms of the recovered and calculated $L$ showed peaks at $\sim$2 and $\sim$4\%. The recovered $L$ values were then used to recover the radii of 281 MS, 40 subgiants, 19 giants, and one pre-main-sequence star (total of 341). The peak of the predicted $R$ errors is found at 2\%, which is equivalent to the peak of the published $R$ errors. 

IAU 2015 GAR B2 opened up a new platform to study the standard bolometric corrections and standard luminosities of stars, which is still virgin to be explored. Thus, we encourage new studies producing more and more standard {\it BC} for all the bands of all photometric systems. Studies like estimating {\it BC} of a star from its spectrum seem to be even more promising also to get interstellar extinction up to the distance of the star from its spectra independently of the other methods which could be used to produce a 3D dust map of the solar neighbourhood to confirm existing ones independently.     

\section{Conclusions}
\begin{enumerate}
\item{With the same defining formula first suggested by \citet{Kuiper1938a}, $ BC_{\xi}=M_{\rm bol} - M_{\xi} = m_{\rm bol} - \xi$, the old verbal definition of {\it BC} used in the first normal science period must change to:  {\it BC} of a star is the difference between its bolometric and one of its filtered (passband) magnitudes without ignoring both of the zero-point constants associated with bolometric and filtered quantities in the absolute scale, unlike the other colours operating in the relative scale ($U-B=B-V=V-R=...=0$ of A0V star), which were set to be zero artificially for a hypothetical star A0V.}

\item{It is not the {\it BC} of a star but $BC-C_{2}$ is always negative, where $C_{2}$ is the zero-point constant of the {\it BC} scale in question.}

\item{The zero-point constants of {\it BC} scales are as many as the number of passbands. $C_{2}(\xi) = C_{\rm{Bol}} - C_{\xi} = c_{\rm{Bol}} - c_{\xi}$, where $\xi$ represents one of the passbands of various photometric systems and capital $C$ implies absolute, while small $c$ for apparent.}

\item{The zero-point constant of a {\it BC} scale is not arbitrary. It is a definite number in magnitude scale unique to the passband.}

\item{The zero-point constant ($C_{2}$) of a {\it BC} scale could be a positive number. This means all $BC<C_2$ for a photometric band are valid, that is, bolometric magnitudes ($M_{\rm bol}$ or $m_{\rm bol}$) of some stars could be dimmer than their filtered magnitudes ($M_{\xi}$, $\xi$). This, however, does not mean the part of the luminosity corresponding to $M_{\xi}$ is greater than the total luminosity. This is because $L_{\xi}$=$L \times 10^{\frac{BC_{\xi} - C_{2}(\xi)}{2.5}}$ is valid to produce $L_{\xi}<L$ for all photometric bands.}

\item{The normal science period (Figure \ref{fig:04}) was ended in 2015 in the year when IAU 2015 GAR B2 issued and the silent revolution \citep{Eker2022} started because all of the problems associated with the three paradigms 1) {\it BC} of a star must always be negative, 2) Bolometric magnitude of a star ought to be brighter than its $V$ magnitude, 3) The zero point of {\it BC} scale is arbitrary) were shown to be resolved.}

\item{The new normal science period will start after the new definition of {\it BC} and consequences itemized above are digested among contemporary astrophysicists.}

\end{enumerate}

\medskip

\section*{Acknowledgements}
We would like to thank Seval Taşdemir and Deniz Cennet Çınar for their contribution in the preparation of the manuscript. This research has made use of NASA’s Astrophysics Data System. 


\spacebref{-10pt}{-30pt}
\bibliographystyle{mnras}
\bibliography{refs}

\bsp	
\label{lastpage}
\end{document}